\documentclass[draftcls,onecolumn,12pt]{IEEEtran}
\usepackage{amssymb}
\usepackage{amsmath}
\usepackage{algorithm}
\usepackage[shortlabels]{enumitem}
\usepackage{xcolor}
\usepackage{url}
\usepackage{subcaption}
\usepackage{enumitem}
\usepackage{algorithmic}
\usepackage{multirow}
\usepackage{multicol}
\usepackage{graphicx}
\usepackage{booktabs}
\usepackage[compress,sort,comma,numbers]{natbib}

\newtheorem{prop}{Proposition}

\newtheorem{lem}{Lemma}

\begin{document}
\vspace*{\stretch{1}}
\noindent
\large\textit{This work has been submitted to the IEEE for possible publication. Copyright may be transferred without notice, after which this version may no longer be accessible.}
\vspace*{\stretch{1}}

\title{QoS-aware Scheduling in 5G Wireless Base Stations}
\author{\IEEEauthorblockN{Reshma Prasad and Albert Sunny, \textit{Member IEEE}} 
		\thanks{The authors are with the Department of Computer Science and Engineering, Indian Institute of Technology Palakkad, India (e-mail: 111914006@smail.iitpkd.ac.in; albert@iitpkd.ac.in).}
	}
	\date{}
\maketitle

\begin{abstract}
5G and beyond networks are expected to support flows with varied \emph{Quality-of-Service (QoS)} requirements under unpredictable traffic conditions. Consequently, designing policies ensuring optimal system utilization in such networks is challenging. Given this, we formulate a long-term time-averaged scheduling problem that minimizes a weighted function of packets dropped by the 5G wireless base station. We then present two policies for this problem. The first is a delay-guaranteed near-optimal policy, and the second is a delay-guaranteed sub-optimal policy that provides flow isolation. We perform extensive simulations to understand the performance of these policies. Further, we study these policies in the presence of a closed-loop flow rate-control mechanism.
\end{abstract}


\section{Introduction}

Network slicing in 5G is a virtualization technique that uses the same physical infrastructure to support many flows. Flows in 5G networks are differentiated in terms of service demands and \emph{Quality-of-Sevice (QoS)} requirements \cite{redana_simone_2019_3265031}. \emph{User Plane Function (UPF)} maps data traffic from the core network to particular QoS flows and forwards them to 5G \emph{Radio Access Network (RAN)} \cite{3GPP}, which buffers data at different points to achieve the desired QoS \cite{irazabal2020dynamic}. One such point of interest is gNodeB --- a 3GPP-compliant implementation of the 5G-NR wireless base station \cite{rinaldi20215g}.

gNodeB is responsible for admission control and scheduling decisions that ensure flow QoS requirements while achieving optimal resource utilization. Admission control requires deciding the number of packets to drop. Whereas, scheduling involves allocating a limited set of \emph{Physical Resource Blocks (PRBs)} to transmit packets amidst variation in traffic and channel conditions \cite{abouaomar2022federated}. Flows can enjoy complete isolation with a static allocation of a disjoint partition of resources. However, this can lead to sub-optimal resource utilization in scenarios with co-existing bursty and non-bursty flows \cite{navarro2020survey,afolabi2018network}. On the other hand, a dynamic scheme ensures better resource utilization, but care should be taken to prevent network state variations from impacting the QoS of flows \cite{li2020hierarchical}.

\subsection{Related Work \label{sec:related_works}}

Over the years, researchers have studied various QoS-aware packet scheduling and optimization problems in 5G networks. In \cite{karimi2019efficient,karimi2020low}, the authors study resource allocation in 5G RAN for co-existing \emph{Ultra-reliable Low Latency Communications (URLLC)} and \emph{Enhanced Mobile Broadband (eMBB)} flows. The allocation problem is formulated as a throughput maximization problem, and a low-complexity solution is proposed where URLLC packets are prioritized over eMBB ones.  However, this can lead to a buffer overrun for eMBB flows.

A \emph{Reinforcement Learning (RL)} based inter-slice allocation and proportional fairness-based intra-slice resource allocation are proposed in \cite{zhou2021ran}. They assume each slice to be an intelligent agent that competes for resources and exchanges $Q$ values with other agents to make allocation decisions. The lack of a knowledge transfer method in this work can lead to poor generalization capability. Consequently, agents need to be re-trained for each new scenario. Further, allocation in the presence of traffic bursts has not been considered. 

 In \cite{hao2022delay}, the authors consider a delay minimization problem and formulate it as a partially observable Markov decision process. They propose an RL-based technique that allocates resources by monitoring parameters such as spectral efficiency, average rate, and queuing delay. The authors note that their technique does not work well with frequently changing traffic patterns. Other model-free and AI-based solutions also suffer from similar issues \cite{comcsa2018towards,comsa20195mart,gu2021knowledge,bae2021learning}.

Lyapunov optimization is a popular technique to control dynamic systems while ensuring stability and performance guarantees. For example, a resource-usage minimization problem with average rate and delay constraints can be solved using this technique \cite{papa2019optimizing}. In \cite{kasgari2018stochastic}, the authors propose a Lyapunov optimization framework for low-latency RAN slicing by considering a power minimization problem with slice isolation and latency violation constraints. A similar approach has been used for joint resource allocation and transmission power minimization of eMBB and URLCC slices in vehicular networks \cite{chen2020network}. Though these works consider several QoS aspects, they do not consider packet loss and the associated QoS degradation. When packet drops are not considered, the Lyapunov optimization framework makes sense only within the capacity region \cite{neely2010stochastic}.

In \cite{jung2020intelligent}, the authors consider a scenario where the controller has no control over arrival and transmission rates. They use \emph{Active Queue Management (AQM)} in conjunction with Lyapunov optimization to minimize packet loss subject to queue stability. Minimizing packet drops leads to an increase in delay, and a workaround for this problem is not presented by the authors. Further, they do not provide any throughput guarantees to individual flows. Opportunistic scheduling, proposed in \cite{neely2011opportunistic}, has been used for dynamic airtime allocation to maximize throughput while adhering to strict delay constraints \cite{richart2020slicing}. Their method requires estimating the maximum arrival rate, and the quality of the estimate directly impacts the QoS.

\subsection{Our Contributions \label{sec:contributions}}
In this paper, we explore policies that guarantee a minimum average service rate and bounded worst-case delay. The main contributions of this paper are as follows:
\begin{itemize}
    \item We formulate a resource allocation problem to minimize a weighted long-term time average of packet drop decisions subject to a guarantee on average service rate and queue stability constraints.
    \item We present two policies: the first one is a delay-guaranteed near-optimal admission and scheduling policy whose performance can be controlled with a couple of parameters, and the second one is a delay-guaranteed sub-optimal policy that provides flow isolation. 
    \item We perform extensive simulations to validate the performance of these policies. Further, we study these policies in the presence of a closed-loop flow rate-control mechanism.  
\end{itemize}

The remainder of the paper is organized as follows. In Sec.~\ref{sec:system_model}, we discuss the system model and problem formulation. Using virtual queues and Lyapunov optimization,  we obtain a delay-guaranteed near-optimal policy in Sec.~\ref{sec:solution}. In Sec.~\ref{sec:static}, we present a delay-guaranteed sub-optimal policy that provides flow isolation. Simulation results and their discussions are presented in Sec.~\ref{sec:simulation}. In Sec.~\ref{sec:congestion}, we study the impact of closed-loop flow rate control on our policies. Finally, in Sec.~\ref{sec:conclude}, we conclude the paper.

\section{System Model and Problem Formulation \label{sec:system_model}}

Consider a scenario where packets from the core network, segregated into a set of $\mathcal{N} = \{1,2,\ldots,n\}$ QoS flows by UPF, are forwarded to a gNodeB. With each flow $i \in \mathcal{N}$, we associate a data queue $\mathcal{Q}_i$ that stores a backlog of packets. Packets in the data queues can be dropped to maintain flows' QoS. We consider a slotted system where the gNodeB makes resource allocation and packet drop decisions at the beginning of each slot.

Let $A_i(t) \in \{0,1,\ldots, A^{max}_i \}$ denote the number of packets that arrive into queue $\mathcal{Q}_i$ in slot $t$. 
 Let $D_i(t) \in \{0,1,\ldots, D^{max}_i\}$ be the drop decision for flow $i$ in the $t^{\textrm{th}}$ slot. Then, the evolution of flow $i$'s queue length is governed by the following recursive equation
\begin{equation} Q_i(t+1)=[Q_i(t)-S_i(t)-D_i(t)]^++A_i(t) \label{eq:dq} \end{equation}
where $[x]^+=max\{0,x\}$, $Q_i(t)$ denotes flow $i$'s queue length (number of packets) at the beginning of slot $t$, and $S_i(t) = f(R_i(t),t) \in \mathbb{Z}^{+}$ is the number of flow $i$'s packets that can be successfully transmitted in the $t^{\textrm{th}}$ slot. 
$f(\cdot,\cdot)$ depends on the channel conditions \cite{neely2011opportunistic}, and is an increasing linear function of $R_i(t)$ --- the number of \emph{Physical Resource Blocks (PRBs)} allocated to flow $i$ in the $t^{\textrm{th}}$ slot \cite{alsenwi2019chance,rinaldi20215g}. As in \cite{neely2010stochastic}, we assume that channel conditions are constant for a slot duration, and gNodeB observes them at the beginning of each slot.
Consequently, the maximum number of packets that can be transmitted in a slot, i.e., $\sum^{n}_{i=1} S_i(t)$, is bounded above by $S(t)=f(R^{total},t) \in [0, 1, \ldots, S^{max}]$, where $R^{total}$ is the total number of PRBs and $S^{max}$ is the maximum attainable transmission rate in any slot. 

\noindent
\textbf{Remark}: \textit{In Eq.~\eqref{eq:dq}, we apply the transmission decision before the drop decision. While this order does not affect the queue evolution, in practice, it will ensure that only surplus packets that cannot be transmitted are dropped.}

\noindent
\textbf{Remark}: \textit{As in \cite{neely2011opportunistic}, we have decoupled the transmission ($S_i(t)$) and drop ($D_i(t)$) decisions from queue backlog $Q_i(t)$ to obtain a dynamic policy that depends only on the current system state. These decisions give an upper bound on the actual number of packets dropped and transmitted. In fact, the actual amount of flow $i$'s packets transmitted in slot $t$ is $\tilde{S}_i(t) = \min \{Q_i(t), S_i(t) \}$. Whereas the actual amount of packets dropped is $\min \{ [Q_i(t)-\tilde{S}_i(t)]^{+},D_i(t) \}$. A comparison of the drop decisions and actual packet drops is presented in Sec.~\ref{sec:simulation_pc}. } 

For each 5G flow, QoS characteristics are specified as parameters associated with \textit{5G QoS Identifier (5QI)} \cite{3GPP}. One such parameter of importance is \textit{Guaranteed Flow Bit Rate (GFBR)} --- the average bit rate guaranteed to be provided to the flow. To accommodate such a requirement, we impose the following long-term time-averaged constraint 
\begin{align}
\liminf_{T \to \infty} \dfrac{1}{T}\sum^{T}_{t=1}  [S_i(t)-\alpha_i S(t)] \geq 0  \label{eq:service_constraint}
\end{align}
\noindent
where $\alpha_i \in [0,1]$. Let ${S}_{avg} =\liminf_{T \to \infty} \frac{1}{T}\sum^{T}_{t=1}S(t)$. Then, due to Constraint~\eqref{eq:service_constraint}, we have
\begin{align*}
\liminf_{T \to \infty} \frac{1}{T} \sum^T_{t=1} S_i(t)
\geq& \liminf_{T \to \infty} \frac{1}{T} \sum^T_{t=1} [S_i(t)-\alpha_i S(t)]\\&+\alpha_i \liminf_{T \to \infty} \frac{1}{T} \sum^T_{t=1} S(t)
\overset{a}{\geq}  \alpha_i {S}_{avg}
\end{align*}
i.e., Constraint~\eqref{eq:service_constraint} ensures that the long-term average service rate is at-least $\alpha_i {S}_{avg}$. For schedule feasibility, the aggregate service rate of flows cannot exceed the maximum available transmission capacity of gNodeB in a slot, i.e., $\sum_{i=1}^{n}\alpha_i \leq 1$. We note that such a constraint on $\alpha_i$'s can be used by gNodeB as an admission criterion for flows with GFBR requirements.

Due to the minimum service rate constraint, the scheduler may need to drop packets to prevent data queues from blowing up. Frequent packet drops at the gNodeB lead to packet re-transmissions by the source resulting in poor end-to-end delays, and wastage of core network bandwidth. With this in mind, we aim to minimize a weighted long-term time average of packet drop decisions. Our optimization problem can be formally stated as follows
\begin{equation}
\tag{P1}
\label{eq:p1}
\begin{aligned}
&  \hspace{5mm}  \min_{\{ S_i(t),D_i(t), i \in \mathcal{N}, t \geq 1 \} } \limsup_{T \to \infty} \dfrac{1}{T}\sum^{T}_{t=1} \sum^n_{i=1} w_i D_i(t) \nonumber \\
&  \textrm{Subject to: }  \\
 & \hspace{5mm} \liminf_{T \to \infty} \dfrac{1}{T}\sum^{T}_{t=1}  [S_i(t)-\alpha_i S(t)] \geq  0 \quad \forall i \in \mathcal{N} \\
 & \hspace{20mm} 0\leq \sum_{i=1}^{n} S_{i}(t) \leq  S(t)  \quad \forall t \geq 0  \\
& \hspace{10mm} S_i(t) \in \mathbb{Z}^{+}  \textrm{ and } D_{i}(t) \in \{0,1,...,D^{max}_i\} \\
 & \hspace{20mm} \textrm{All data queues are rate stable} 
\end{aligned}
\end{equation}
where $w_i \in [0,1]$ is the weight assigned to flow $i$'s drop decision.

The above optimization framework can also be used for 5G intra-slice resource allocation, i.e., allocating packet service rate to flows with the same slice type. In such a setting,  $S(t)$ would correspond to the service rate allocated to the slice type, and $\alpha_i$'s would correspond to the fine-grained QoS requirement of flows belonging to this slice type.

\section{A Delay-Guaranteed Near-Optimal Policy \label{sec:solution}}

Problem~\eqref{eq:p1} needs to be solved considering the stability of data queues and adherence to the long-term time-averaged constraint on data packet service rate. In this section, by constructing virtual queues and using the \emph{Lyapunov drift-penalty} technique, we reduce this problem to a series of optimization problems that can be solved in each time slot. We show that the policy thus obtained is near-optimal with delay guarantees.

\subsection{Virtual Queues \label{sec:virtual_queuees}}
We handle Constraint~\eqref{eq:service_constraint} with virtual queues that transform the long-term time-averaged inequality constraint into a queue stability problem \cite{neely2011opportunistic,neely2010stochastic}. Consequently, any algorithm stabilizing the virtual queues satisfies the long-term time-averaged constraint. For flow $i \in \mathcal{N}$, we consider a virtual queue $\mathcal{Y}_i$ with a queue length described by the following Lindley equation.
  \begin{equation}
\begin{aligned}
Y_i(t+1)=[Y_i(t)  +\alpha_i S(t)-S_i(t)]^+ \label{eq:yq} 
 \end{aligned}
\end{equation}
with $Y_i(0) = 0$. The following proposition shows that Constraint~\eqref{eq:service_constraint} is satisfied if the above-defined virtual queues are rate stable.

\begin{prop} \label{prop:serv_const}
If each of the virtual queues $\mathcal{Y}_i$ are rate stable, i.e., $\limsup_{T \to \infty} \dfrac{Y_i(T)}{T}=0 \quad \forall i \in \mathcal{N}$, then Constraint~\eqref{eq:service_constraint} is satisfied.
\end{prop}
\begin{IEEEproof}
Refer Appendix~\ref{sec:serv_const_proof}.
\end{IEEEproof}

\subsection{Persistent Queues \label{sec:persistent_queuees}}

In addition to a minimum average service rate requirement, flows often have an upper bound on end-to-end delay. Therefore, it is pertinent to quantify/bound delay experienced by flows at gNodeB. While the optimal solution to the problem presented in Sec.~\ref{sec:system_model} ensures that data queues do not blow up, the length of these queues can be arbitrarily large. To address this issue, we use persistent virtual queues \cite{neely2011opportunistic}. For flow $i \in \mathcal{N}$, we consider a persistent virtual queue $\mathcal{Z}_i$ whose length evolves as per the following recursive equation.
\begin{equation} 
Z_i(t+1)=[Z_i(t) + \zeta \cdot (\alpha_i S(t) \mathbb{I}_i(t) - S_i(t)- D_i(t))]^+ \label{eq:zq}
\end{equation}

Here, $\mathbb{I}_i(t) \in \{0,1 \}$ is an indicator function that takes the value 1 if and only if $Q_i(t)>0$, and $\zeta > 0$ is a parameter that determines the trade-off between delay and deviation from optimality. We choose $Z_i(0) = 0$. When flow $i$'s data queue is not served despite having packets to transmit, $Z_i(t)$ increases at a rate of $\zeta \alpha_i S(t)$. Stable persistent queues ensure that packet drop/transmit decisions are made within a finite time. An upper bound on this decision time is presented in Proposition~\ref{prop:delay_dynamic}.

\noindent 
\textbf{Remark: } \textit{For ease of presentation, the value of the parameter $\zeta$ is same for all flows. Our approach works even for an individualized set of parameters $\{\zeta_i, i \in \mathcal{N}\}$. From Proposition~\ref{prop:delay_dynamic}, one can see that such individualization allows for fine-grained control of delay experienced by flows.}

\subsection{Lyapunov Optimization \label{sec:lyapunov}}

Let us define the following quadratic function
$$L(t)=\frac{1}{2} \sum_{i=1}^n (Z_i(t)^2+Q_i(t)^2+Y_i(t)^2)$$

The Lyapunov drift is defined as $\Delta L(t)=L(t+1)-L(t)$. Minimizing the drift keeps the length of the queues finite. However, it can lead to large packet drops. Consequently, as in \cite{neely2010stochastic}, we minimize an upper bound on the drift-plus-penalty expression $\Delta L(t)+V D(t)$, where  $V\geq0$ is a parameter that controls the trade-off between optimality gap and convergence rate, and $D(t)=\sum_{i=1}^n w_i D_i(t) $ is the penalty term. The drift-plus-penalty function can be bounded above as follows (refer Appendix~\ref{sec:drift} for the derivation).
\begin{align}
&L(t+1)-L(t)+V D(t) \leq  C_1 + \zeta^2 C_2 \nonumber \\
& \hspace{5mm} + \sum^n_{i=1} \zeta Z_i(t)[\alpha_i  S(t) \mathbb{I}_{i}(t)-S_i(t)-D_i(t)] \nonumber \\
& \hspace{5mm} + \sum^n_{i=1} Q_i(t)[A_i(t)-S_i(t)-D_i(t)] \nonumber \\
& \hspace{5mm} + \sum^n_{i=1} Y_i(t)[\alpha_i S(t)-S_i(t)]+V \sum_{i=1}^{n}w_i D_i(t) \label{eq:drift_p}
\end{align}
where $C_1 = n (S^{max} )^2 + \sum^n_{i=1} (A^{max}_i + S^{max} + D^{max}_i)^2/2 $ and $C_2 = \sum^n_{i=1} (S^{max} + D^{max}_i)^2$.

Now, to obtain a control algorithm for slot $t$, we minimize the RHS of Inequality~\eqref{eq:drift_p} with respect to decision variables $\{(S_i(t), D_i(t)), i \in \mathcal{N}\}$.  This minimization problem can be decomposed into two independent sub-problems. One of them is the \emph{Service Rate Allocation (SRA) sub-problem} that can be stated as follows
\begin{align*}
&\hspace{5mm} \max_{\{ S_i(t), i \in \mathcal{N} \} } \sum_{i=1}^{n} [\zeta Z_i(t)+Q_i(t)+Y_i(t)]S_i(t)\\
& \hspace{-3mm} \textrm{Subject to: } 
 0\leq \sum_{n=1}^{n} S_{i}(t) \leq  S(t)\leq S^{max} \textrm{ and } S_i(t) \in \mathbb{Z}^{+}
 \end{align*}
 
The above problem aims to maximize a convex combination of non-negative decision variables. Consequently, the maxima can be achieved by providing the entire service $S(t)$ to the flow with the largest value of $(\zeta Z_i(t)+Q_i(t)+Y_i(t))$, i.e., 
\begin{equation}
\overline{S}_i(t) = \begin{cases}
S(t) & \textrm{if } i = \overline{i}(t) \\
0 & \textrm{otherwise}
\end{cases}
 \label{eq:p2_txdecision}
\end{equation}
where $\overline{i}(t) = \arg \max_{k \in \mathcal{N} } [\zeta Z_k(t)+Q_k(t)+Y_k(t)]$. Tie, if any, can be broken using any arbitrary rule.

The second sub-problem is the \emph{Drop Decision (DD) sub-problem} that has the following form. \begin{align*}
&\max_{\{ D_i(t),  i \in \mathcal{N} \} } \sum_{i=1}^{n} [\zeta Z_i(t)+Q_i(t)-V w_i]D_i(t)\\
& \hspace{7mm} \textrm{Subject to: } D_i(t) \in \{ 0, 1, \ldots, D^{max}_i\}
 \end{align*}
 
The optimal drop decision has a threshold structure where the decision is to drop packets only when the aggregate weighted-queue length $Q_i(t)+\zeta Z_i(t)$ exceeds $Vw_i$, i.e., 
 \begin{equation}
 \label{eq:p2_dropdecision}
 \overline{D}_i(t)=\begin{cases}
 D^{max}_i & \textrm{ if } Q_i(t)+\zeta Z_i(t)>V w_i  \\
0 & \textrm{otherwise} \end{cases} \end{equation}

Choosing $D^{max}_{i} = 0$ leads to a long-term time-averaged weighted aggregate packet drop of zero --- the least possible value. However, such a choice may not ensure the stability of data queues. A natural question at this point is: how should $D^{max}_{i}$ be chosen so that \eqref{eq:p1} has a feasible solution? The following proposition addresses this question. Let $\boldsymbol{\overline{\pi}}=\{(\overline{S}_i(t),\overline{D}_i(t)), i \in \mathcal{N}\}^{\infty}_{t=1}$ be the policy obtained from decision rules \eqref{eq:p2_txdecision} and \eqref{eq:p2_dropdecision}. 

\begin{prop} \label{prop:pio_feasible}
If, for all $i \in \mathcal{N}$, $D^{max}_i$ is at least $max\{A^{max}_i,\alpha_i S^{max}\}$, then policy $\boldsymbol{\overline{\pi}}$ is a feasible solution of \eqref{eq:p1}.
\end{prop}
\begin{IEEEproof}
Refer Appendix~\ref{sec:pio_feasible_proof}.
\end{IEEEproof}

While policy $\boldsymbol{\overline{\pi}}$ is a feasible solution of \eqref{eq:p1}, computing it requires knowledge of the maximum number of packet arrivals in a slot ($A^{max}_i$). In a real-world setting, this upper bound may not be known prior due to the difficulty in precisely characterizing flows' packet arrival process. Therefore, we propose the following drop decision that only requires knowledge of packet arrivals in the current slot.
\begin{equation}
\hat{D}_i(t)= \begin{cases}
  \max\{A_i(t),\alpha_i S(t)\} & \textrm{if } Q_i(t)+\zeta Z_i(t)>V w_i  \\
0 & \textrm{otherwise} \end{cases} \label{eq:p2_dropdecision_2}
\end{equation}

Let $\boldsymbol{\hat{\pi}}=\{(\overline{S}_i(t),\hat{D}_i(t)), i \in \mathcal{N}\}^{\infty}_{t=1}$ be the policy obtained from decision rules \eqref{eq:p2_txdecision} and \eqref{eq:p2_dropdecision_2}. 

\begin{prop} \label{prop:pih_feasible}
Policy $\boldsymbol{\hat{\pi}}$ is a feasible solution of \eqref{eq:p1}.
\end{prop}
\begin{IEEEproof}
Similar to the proof of Proposition~\ref{prop:pio_feasible}.
\end{IEEEproof}

Let ${D}^*$ be the long-term time-averaged weighted drops obtained from an optimal solution $\boldsymbol{\pi^{*}} =\{(S_i^*(t),D_i^*(t)) , i \in \mathcal{N} \}^{\infty}_{t=1}$ of \eqref{eq:p1}. The following proposition shows that policy $\boldsymbol{\hat{\pi}}$ is near-optimal, i.e., it can achieve long-term time-averaged weighted drops arbitrarily close to the optimal value.

\begin{prop} \label{prop:performance}
For any $\epsilon > 0$, $|{D}^* - \hat{D}| \leq O( \epsilon)$ for a large enough value of parameter 
 $V$. Here, $O(\epsilon)$ represents a positive quantity linearly going to 0 as $\epsilon$ goes to zero, and $\hat{D}=\limsup_{T \to \infty} \frac{1}{T} \sum^T_{t=1} \sum^n_{i=1} w_i \hat{D}_i(t)$.
\end{prop}
\begin{IEEEproof}
Refer Appendix~\ref{sec:performance_proof}.
\end{IEEEproof}

\subsection{Worst-case Delay} 
Lemma~\ref{lem:qbounds_under_hat} establishes that data queue lengths are bounded under policy $\boldsymbol{\hat{\pi}}$. However, it does not provide any insight into the delay experienced by packets. In fact, for the problem considered in the paper, it is possible to construct policies that result in bounded queue lengths and unbounded packet delays. We recollect that persistent queues were included in the \emph{Lyapunov function} to bound queuing delays. 

\begin{prop}\label{prop:delay_dynamic}
If $S^{max} \geq S(t) \geq S^{min} > 0$, then under policy $\boldsymbol{\hat{\pi}}$, the worst-case delay (no. of slots) of flow $i$'s packets at gNodeB is at most $\frac{S^{max}}{S^{min}}  + \left(1+\frac{1}{\zeta^2}\right) \frac{Vw_i}{\alpha_i S^{min}}+\frac{A_i^{max}}{\alpha_i S^{min}}$. 
\end{prop}
\begin{IEEEproof}
Refer Appendix~\ref{sec:delay_dynamic_proof}.
\end{IEEEproof}

The bound presented in the above proposition can be loose because packets can get dropped well before $Q_i(t)$ or $Z_i(t)$ reaches the bound in Lemma~\ref{lem:qbounds_under_hat}. This happens because drop decisions are not based on individual queue lengths but on the weighted queue length $Q_i(t)+\zeta Z_i(t)$. We note that worst-case delay has a negative correlation with parameter $\zeta$. Therefore, choosing a large value of $\zeta$ can result in lower delays. However, to obtain near-optimal policies, the value of the parameter $V$ should be much larger than $\zeta^2$ (refer to the proof of Lemma~\ref{lem:diff_bound}).

\section{Isolation of Flows \label{sec:static}}

While policy $\boldsymbol{\hat{\pi}}$ is near-optimal, flows remain coupled due to the rate allocation decision \eqref{eq:p2_txdecision}. However, strict flow isolation is desirable in certain scenarios \cite{bosk2021using}. To that end, we consider a policy $\boldsymbol{{\pi}^s}=\{({S}^{s}_i(t),{D}^{s}_i(t)), i \in \mathcal{N}\}^{\infty}_{t=1}$, where $S^{s}_i(t)=\alpha_i S(t) \, \forall t \geq 1$  and
$$D^{s}_i(t)=\begin{cases}
A_i(t) & Q_i(t)>V w_i  \\
0 & \textrm{else} \end{cases}$$

The term $\zeta Z_i(t)$ does not appear in the above decision rule because the constant service rate of $\alpha_i S(t)$ forces $Z_i(t) = 0 \, \forall t \geq 1$. Policy $\boldsymbol{{\pi}^{s}}$ is a feasible solution of \eqref{eq:p1}, but it may not be near-optimal. Nevertheless, as shown in the following proposition, this policy does provide delay guarantees.

\begin{prop} \label{prop:delay_static}
If $S(t) \geq S^{min} > 0$, under policy $\boldsymbol{{\pi}^{s}}$, the worst-case delay (no. of slots) of flow $i$'s packets at gNodeB is bounded above by $1 +  {Vw_i}/{\alpha_i S^{min}}+{A_i^{max}}/{\alpha_i S^{min}}$.
\end{prop}
\begin{IEEEproof}
From the proof of Lemma~\ref{lem:data_queue_bound}, it is easy to see that flow $i$'s queue length is bounded above by  $Vw_i+A_i^{max}$ under policy $\boldsymbol{{\pi}^{s}}$. Flow $i$ has a  packet service rate of $\alpha_i S(t)$ that is bounded below by $\alpha_i S^{min}$, and transmission decisions are made at the beginning of each slot. Consequently, the worst-case delay (no. of slots) of a packet arriving in a slot is bounded above by $1 + \lfloor (Vw_i+A_i^{max}) / \alpha_i S^{min} \rfloor \leq 1 + (V w_i + A_i^{max}) / \alpha_i S^{min}$.
\end{IEEEproof}

\noindent
\textbf{Remark}: \textit{From Propositions~\ref{prop:delay_dynamic} and \ref{prop:delay_static}, one can see that policy
$\boldsymbol{{\pi}^{s}}$ provide a better delay guarantee than policy $\boldsymbol{\hat{\pi}}$. The difference of $O(V/\zeta^2)$ in the guarantees arises due to the dynamic rate allocation mechanism in policy $\boldsymbol{\hat{\pi}}$. }

\section{Simulations \label{sec:simulation}}
In this section, we discuss results from simulations that give insights into the performance of our policies. In all simulations, we consider a total of $10^5$ unit length slots. The maximum packet service rate of gNodeB is taken as $S(t)= 50$ packets $\forall t \geq 1$. $w_i$ is set as $1$ for all flows $i \in \mathcal{N}$. 


Flow $i$'s packet arrival process $\{A_i(t), t \geq 1 \}$ is considered to be a collection of i.i.d. random variables with the following probability mass function\footnote{Form of the mass function is motivated by the well-known bursty traffic model \emph{FTP model 3} \cite{mogensen2021empirical, navarro2020survey}.} 
 \begin{equation*}
    P(A_i(t)=\eta_i k)=
\begin{cases}
\frac{e^{-\lambda_i}\lambda_i^k}{k!} & 0 \leq k \leq \nu_i-1 \\
1-\sum_{l=0}^{\nu_i-1}\frac{e^{-\lambda_i}\lambda_i^l}{l!} &  k=\nu_i\\
0  & k > \nu_i
\end{cases}
\end{equation*}

\begin{table}[!t]
\renewcommand{\arraystretch}{1.3}
    \centering
    \caption{Service requirement and arrival process combinations used for simulations.  }
    \label{tab:parameters}
    \begin{tabular}{ccccc}
     \hline
        
         \textbf{No.} & $\boldsymbol{\alpha_1}$ & $\boldsymbol{\alpha_2}$ & \textbf{Arrival process} & \textbf{Implication} \\
         \hline \hline
         \multirow{2}{*}{1} & \multirow{2}{*}{0.2}  & \multirow{2}{*}{0.8} & $\mathcal{A}_1=(1,10,300)$ & $\sum_{i\in \mathcal{N}} \alpha_i=1$\\
          &   &  & $\mathcal{A}_2=(1,40,300)$ & $\lambda_i=\alpha_i S_{avg} \forall i \in \mathcal{N}$  \\

         \multirow{2}{*}{2} & \multirow{2}{*}{0.2}  & \multirow{2}{*}{0.4} &    $\mathcal{A}_1=(1,30,300)$  & $\sum_{i\in \mathcal{N}} \alpha_i<1$  \\
          &  &  & $\mathcal{A}_2=(1,70,300)$ & $\lambda_i>\alpha_i S_{avg} \forall i \in \mathcal{N}$  \\

         \multirow{2}{*}{3} & \multirow{2}{*}{0.2}  & \multirow{2}{*}{0.6} & $\mathcal{A}_1=(\eta,10/\eta,300/\eta)$  & $\sum_{i\in \mathcal{N}} \alpha_i<1$\\
         & & & $\mathcal{A}_2=(\eta,30/\eta,300/\eta)$ &  $\lambda_i \eta_i =\alpha_i S_{avg} \forall i \in \mathcal{N}$  \\
         \hline
       
    \end{tabular}
\end{table}

 Flow $i$'s packet arrival process is characterized by the 3-tuple $(\eta_i,\lambda_i,\nu_i)$, where $\eta_i \in \mathbb{Z}^{+}$ is the burst size, $\lambda_i \in \mathbb{R}^{+}$ and $\nu_i \in \mathbb{Z}^{+}$. In the remainder of the paper, we will use the above tuple to denote an arrival process. For the above packet arrival process $A^{max}_i = \eta_i \nu_i$, and the average packet arrival rate is $\lambda_i \eta_i$.

The combinations of service requirements and arrival processes used for simulations in this section are presented in Table~~\ref{tab:parameters}. These specific combinations have been selected taking into consideration the following factors: 
\begin{itemize}
    \item When $\sum_{i\in \mathcal{N}} \alpha_i < 1$, the average service rate of flow can be higher than the guaranteed rate since they can utilize the unused air-time. 
    \item When the average arrival rate of a flow is higher than its guaranteed minimum service rate, gNodeB can drop a large number of packets to ensure QoS.
\end{itemize}

\subsection{Effect of Parameter $\zeta$}
 \begin{figure*}[!t]
    \centering
    \begin{subfigure}{0.4\linewidth}
    \centering
    \includegraphics[width=\linewidth]{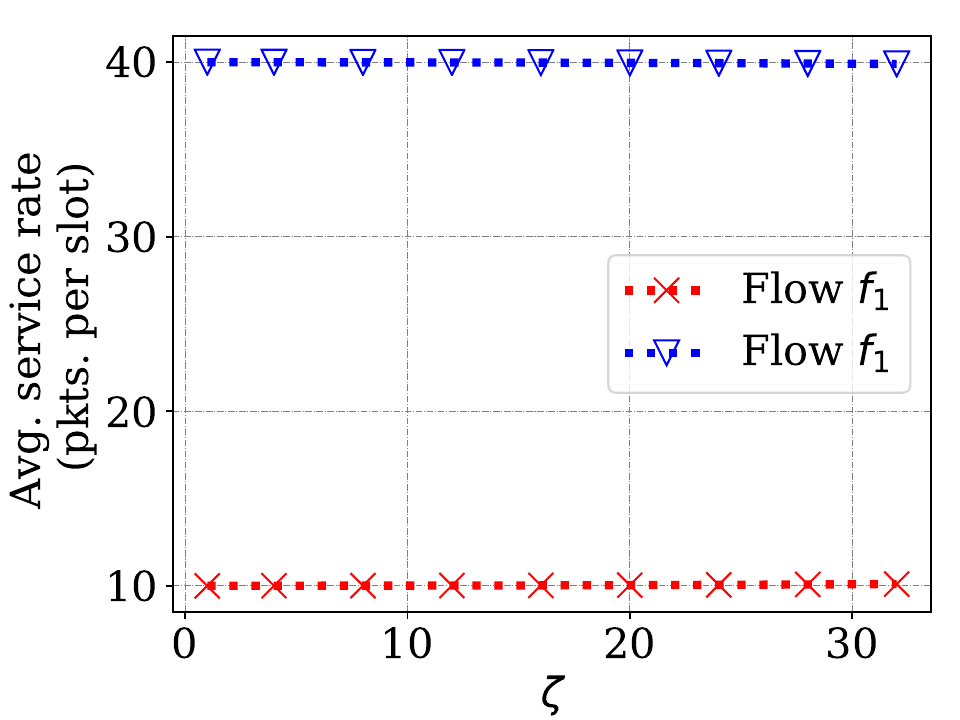}
    \caption{Average service rate} \label{fig:zeta_service}
    \end{subfigure}
     \begin{subfigure}{0.4\linewidth}
     \centering
    \includegraphics[width=\linewidth]{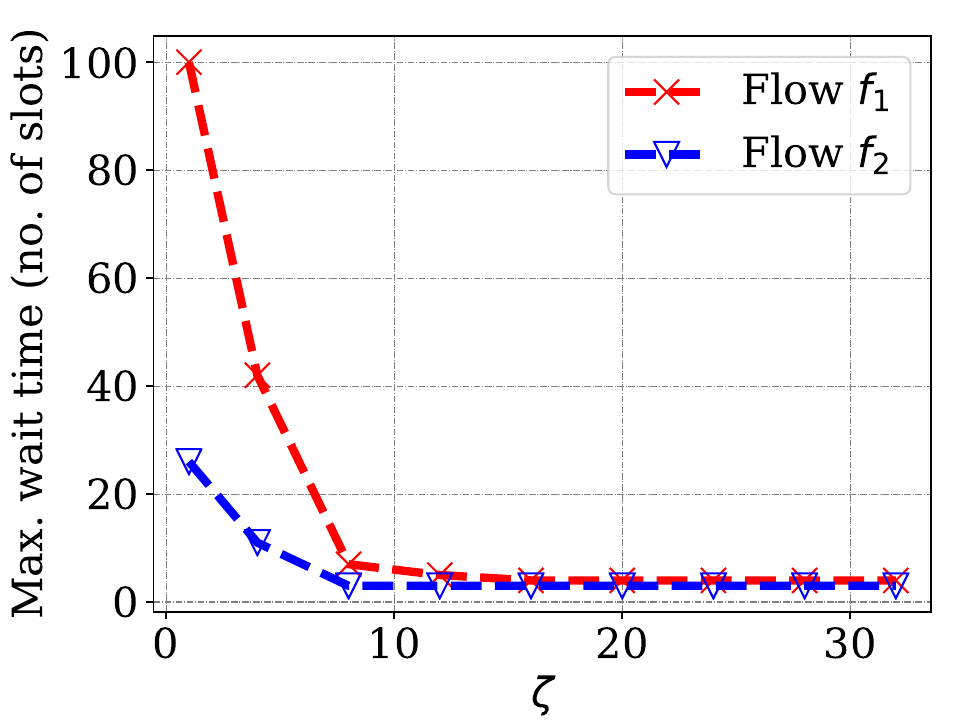}
    \caption{Maximum wait time at gNodeB} \label{fig:zeta_wait} 
    \end{subfigure}
    \begin{subfigure}{0.4\linewidth}
    \centering
    \includegraphics[width=\linewidth]{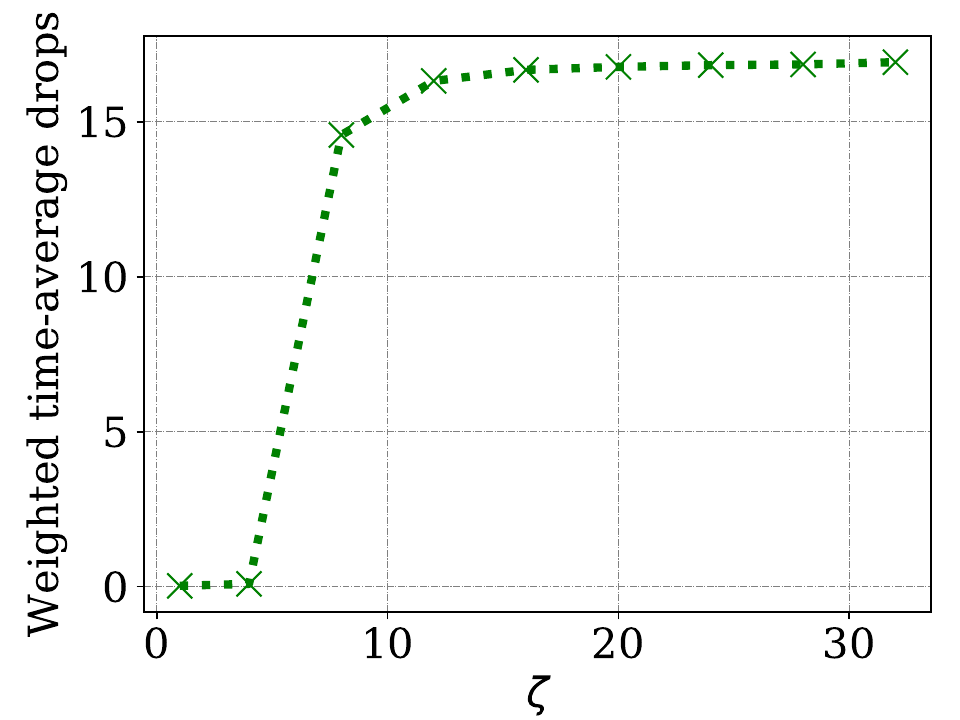}
    \caption{Weighted time-average packet drops} \label{fig:zeta_drop}
    \end{subfigure}
    \caption{Variation of various quantities as a function of parameter $\zeta$ under policy $\boldsymbol{\hat{\pi}}$: $V=1000$; flows $f_1$ and $f_2$ with arrival processes $\mathcal{A}_1=(1,10,300)$ and $\mathcal{A}_2=(1,40,300)$; $\alpha_1 = 0.2$ and $\alpha_2 = 0.8$.} \label{fig:zeta}
\end{figure*}
 We recollect that the evolution of persistent queues is modulated by the parameter $\zeta$. Consequently, we study its impact on the performance of policy $\boldsymbol{\hat{\pi}}$ by fixing the parameter $V$ as 1000. Service requirements and arrival process are as in Combination~1 of Table~\ref{tab:parameters}. Results of the simulation are presented in Fig.~\ref{fig:zeta}. In these figures, the range of $\zeta$ is $[1,31]$ because values larger than $V^{1/2}=1000^{1/2}$ can lead to significant sub-optimality (refer to the proof of Lemma~\ref{lem:diff_bound}).

From Fig.~\ref{fig:zeta_service}, we can see that $\zeta$ does not have a significant effect on the average service rate. On the other hand, there is a drastic reduction in the maximum waiting time as $\zeta$ increases (refer to Fig.~\ref{fig:zeta_wait}). This reduction is in line with the term $O(V/\zeta^2)$ in Prop.~\ref{prop:delay_dynamic}. However, as established in Prop~\ref{prop:performance}, such a reduction comes at the cost of an increase in packet drops as seen in Fig.~\ref{fig:zeta_drop}. This increase mirrors the decrease in the maximum wait time since parameter $\zeta$ enables a trade-off between delay guarantees and deviation from optimality. \textit{In the subsequent sections, $\zeta$ is fixed as 1.}

\subsection{Comparison of Policies $\boldsymbol{\hat{\pi}}$ and $\boldsymbol{\overline{\pi}}$ \label{sec:simulation_pc}} 

\begin{figure}[!t]
    \begin{subfigure}{\linewidth}
    \centering
    \includegraphics[width=0.4\linewidth]{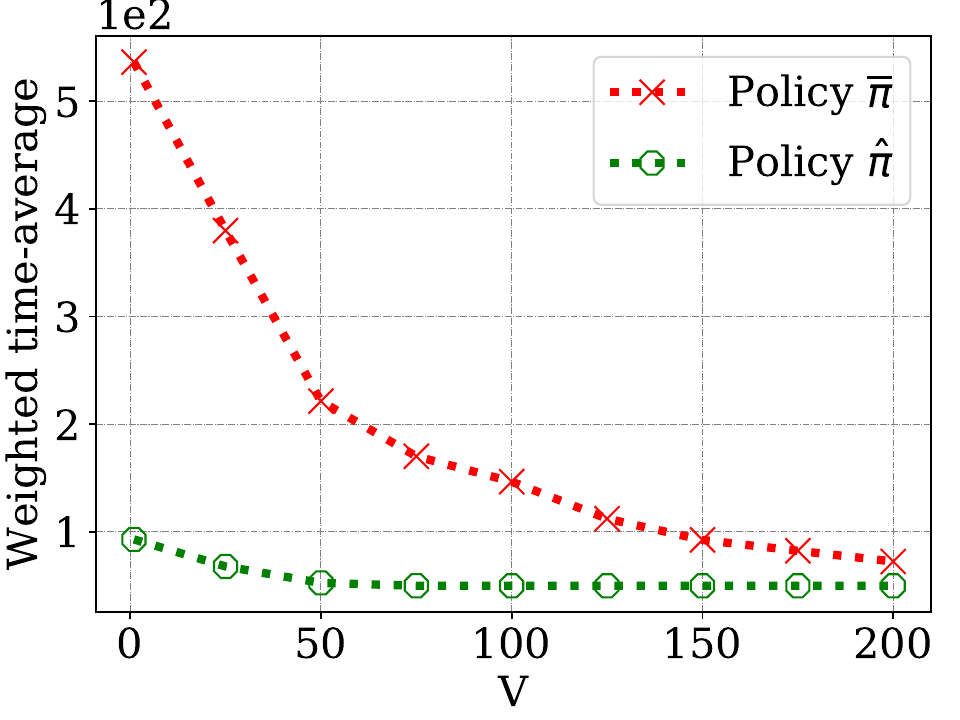}
    \caption{Drop decisions} \label{fig:dover_dhat_drop_decision}
    \end{subfigure}
    \begin{subfigure}{\linewidth}
    \centering
    \includegraphics[width=0.4\linewidth]{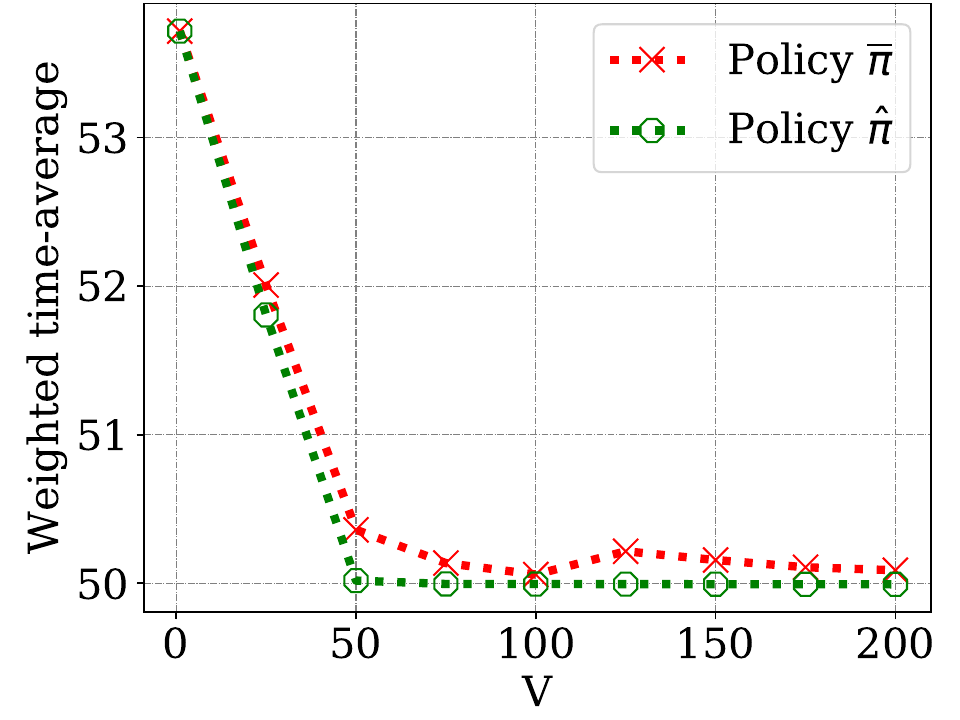}
    \caption{Actual packet drops}\label{fig:dover_dhat_drop_actual}
    \end{subfigure}
    \caption{Weighted time-averaged drops under policies $\boldsymbol{\hat{\pi}}$ and $\boldsymbol{\overline{\pi}}$: 2 flows with arrival processes $\mathcal{A}_1=(1,30,300)$ and $\mathcal{A}_2=(1,70,300)$; $\alpha_1 = 0.2$, $\alpha_2  = 0.4$ and $\zeta=1$. }\label{fig:dover_dhat_drop}
\end{figure}

We consider the service requirements and arrival process as in Combination~2 of Table~\ref{tab:parameters}. Fig.~\ref{fig:dover_dhat_drop_decision} presents the variation in the weighted time-average of drop decisions i.e., $\{D_i(t), i \in \mathcal{N}, t \geq 1  \}$, whereas Fig.~\ref{fig:dover_dhat_drop_actual} presents the variation in the weighted time-average of actual packet drops $\{\tilde{D}_i(t), i \in \mathcal{N}, t \geq 1  \}$. Refer to the second remark in Sec.~\ref{sec:system_model} to know how the actual drops are computed.
 
\begin{table}[!t]
\renewcommand{\arraystretch}{1.3}
    \centering
     \caption{Queue length  statistics under policies $\boldsymbol{\overline{\pi}}$ and $\boldsymbol{\hat{\pi}}$. }\label{tab:qLength}
    \resizebox{0.6\linewidth}{!}{
    \begin{tabular}{ccccccccc}
    \hline
 \multirow{3}{2em}{$\boldsymbol{V}$} & \multicolumn{2}{c}{\textbf{$\boldsymbol{\overline{\pi}}$, flow $f_1$}}&\multicolumn{2}{c}{\textbf{$\boldsymbol{\hat{\pi}}$, flow $f_1$}}&\multicolumn{2}{c}{\textbf{$\boldsymbol{\overline{\pi}}$, flow $f_2$}}&\multicolumn{2}{c}{\textbf{$\boldsymbol{\hat{\pi}}$, flow $f_2$}}\\
  \cmidrule(lr){2-3} \cmidrule(lr){4-5} \cmidrule(lr){6-7} \cmidrule(lr){8-9} 
  &   \textbf{Avg.} & \textbf{Std.} & \textbf{Avg.} &    \textbf{Std} & \textbf{Mean} &    \textbf{Std.}& \textbf{Avg.} &    \textbf{Std} \\
     &   & \textbf{Dev.} &  & \textbf{Dev.} &  & \textbf{Dev.} &  &  \textbf{Dev.} \\
       \hline
       \hline
50 &19.2 &	15.9 & 29.6& 11.7&	13.9 &  15.9 & 15.5 & 16.7  \\
100 & 47.5 & 27.3 &  78.6 &	12.3 & 39.6 & 36.3 & 60.7 & 21.2  \\
150 & 80.8  & 41.1 & 128.6 & 12.3 & 73.6 & 47.6  & 110.7 & 21.2 \\
200 & 111.4 & 54.3 & 178.6 & 12.3 & 98.4 & 62.6 & 160.8 & 21.2 \\
\hline
\end{tabular}}
\end{table}

Smaller values of $V$ lead to frequent drop decisions under both policies. Since the magnitude of drop decisions under policy $\boldsymbol{\overline{\pi}}$ are larger than the ones under policy $\boldsymbol{\hat{\pi}}$, there is a large gap between the plots in Fig. \ref{fig:dover_dhat_drop_decision} for small values of $V$. However, this gap reduces as $V$ increases due to the near-optimal nature of both these policies. While policy $\boldsymbol{\overline{\pi}}$ drops more packets than policy $\boldsymbol{\hat{\pi}}$, the difference in drops scales sub-linearly with time. This leads to the comparable weighted average, of actual packet drops, presented in Fig.~\ref{fig:dover_dhat_drop_actual}. In fact, for large values of $V$, the curves in Figs.~\ref{fig:dover_dhat_drop_decision} and \ref{fig:dover_dhat_drop_actual} converge to the same value.

\begin{figure}[!t]
    \begin{subfigure}{\linewidth}
    \centering
    \includegraphics[width=0.4\linewidth]{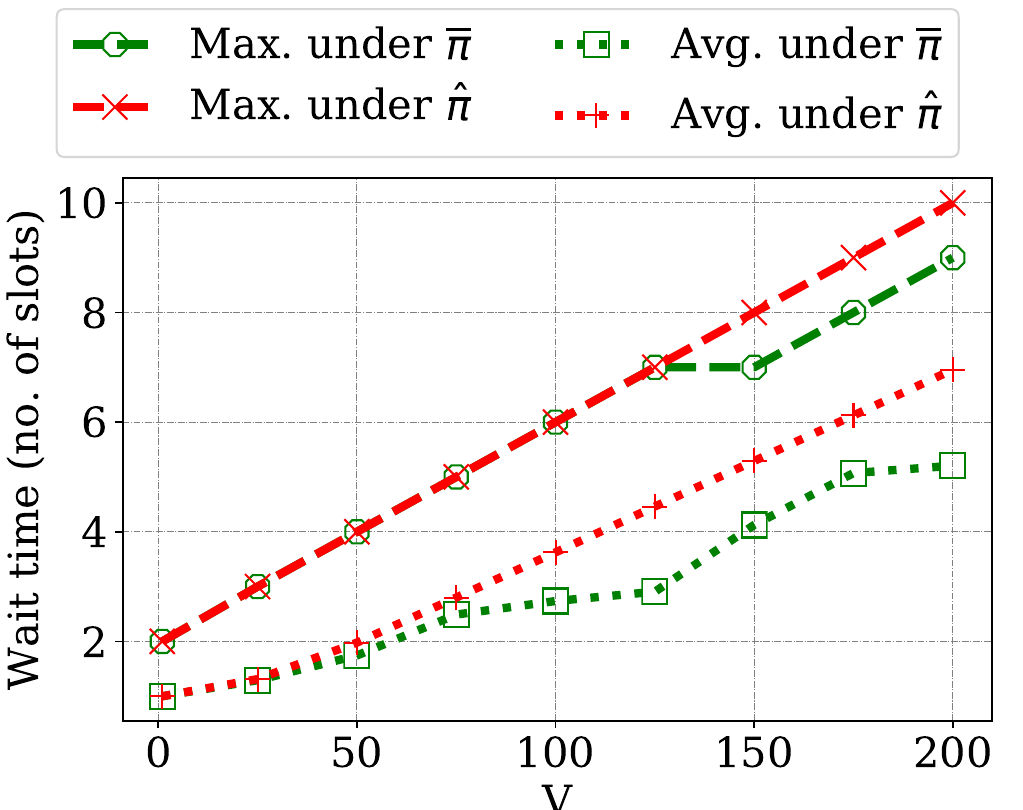}
    \caption{Flow $f_1$}\label{fig:dover_dhat_wait_f1}
    \end{subfigure}
    \begin{subfigure}{\linewidth}
    \centering
    \includegraphics[width=0.4\linewidth]{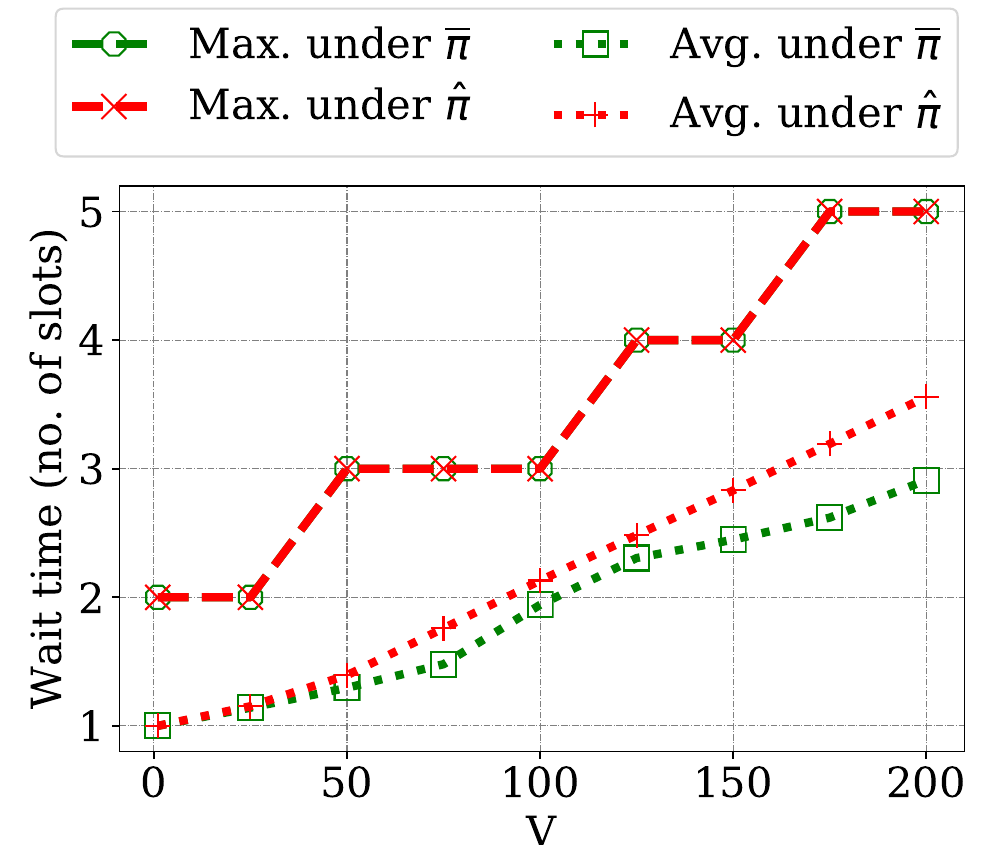}
    \caption{Flow $f_2$} \label{fig:dover_dhat_wait_f2}
    \end{subfigure}
    \caption{Wait time at gNodeB under policies $\boldsymbol{\hat{\pi}}$ and $\boldsymbol{\overline{\pi}}$: 2 flows with arrival processes $\mathcal{A}_1=(1,30,300)$ and $\mathcal{A}_2=(1,70,300)$; $\alpha_1 S^{max} = 0.2$, $\alpha_2 = 0.4$, $V=1000$ and $\zeta=1$.} \label{fig:dover_dhat_wait}
\end{figure}

From Table \ref{tab:qLength}, we observe that the average length of data queues is smaller under policy  $\boldsymbol{\overline{\pi}}$, whereas queue length variation is smaller under policy $\boldsymbol{\hat{\pi}}$. Due to the difference in average queue length, policy $\boldsymbol{\hat{\pi}}$ leads to slightly larger wait times (refer to Figs.~\ref{fig:dover_dhat_wait_f1} and \ref{fig:dover_dhat_wait_f2}). We note that, as established in Props.~\ref{prop:delay_dynamic} and \ref{prop:delay_static}, the maximum wait times grow linearly with $V$ under both policies.

\begin{figure*}[!t]
 \centering
    \begin{subfigure}{0.4\linewidth}
    \centering
    \includegraphics[width=\linewidth]{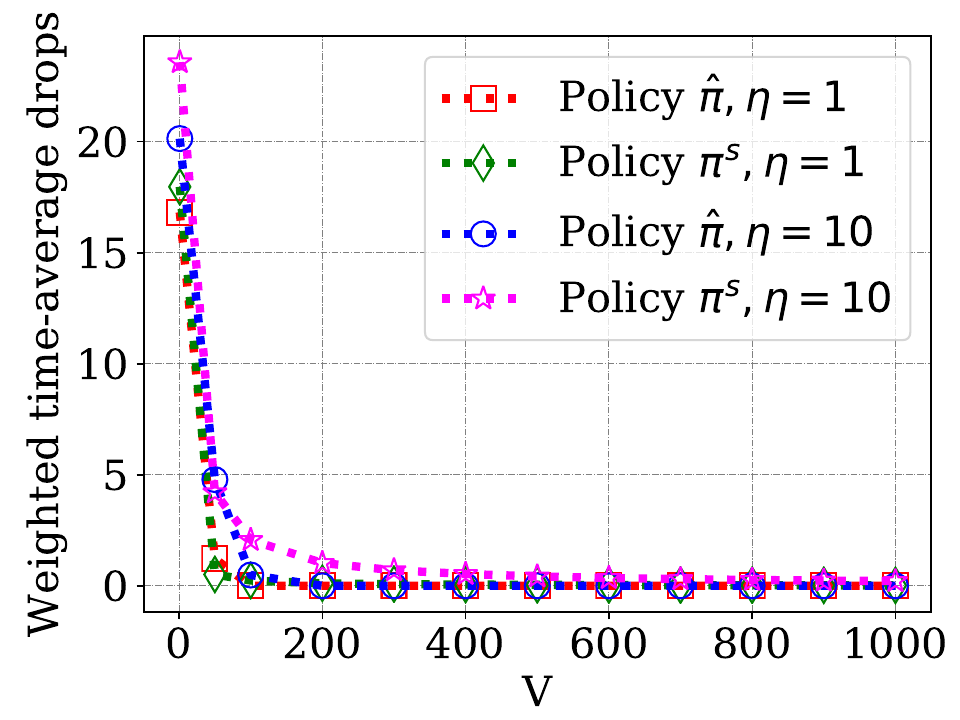}
    \caption{Average weighted drops} \label{fig:bursty_drop}
    \end{subfigure}
    \begin{subfigure}{0.4\linewidth}
    \centering
    \includegraphics[width=\linewidth]{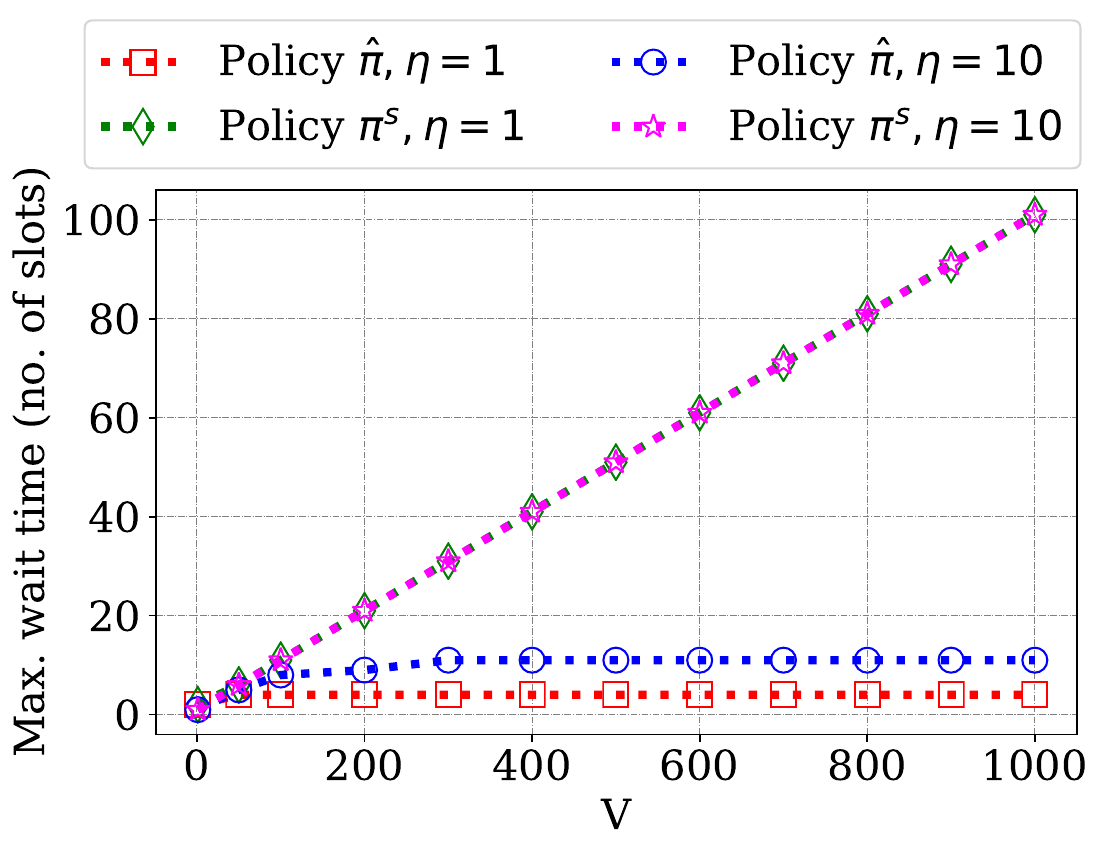}
    \caption{Maximum wait time at gNodeB for flow $f_1$'s packets.}\label{fig:bursty_wait1}
    \end{subfigure}
    \begin{subfigure}{0.4\linewidth}
    \centering
    \includegraphics[width=\linewidth]{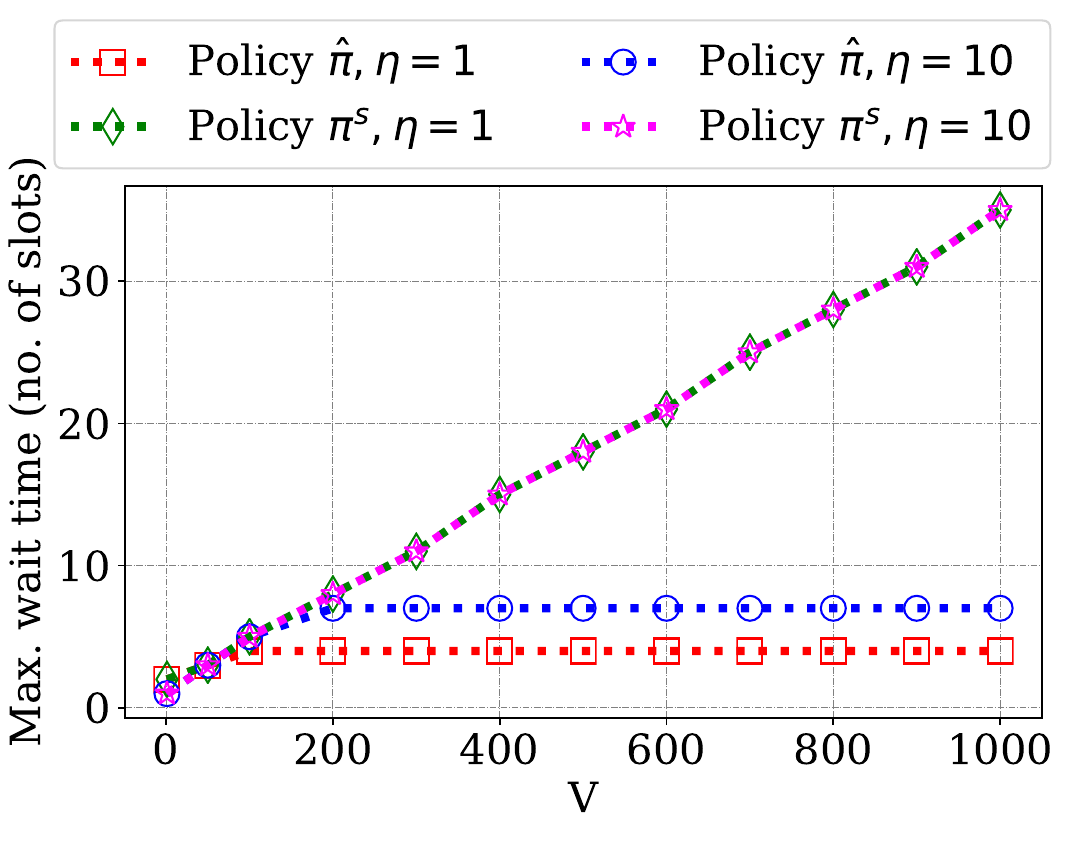}
    \caption{Maximum wait time at gNodeB for flow $f_2$'s packets.}\label{fig:bursty_wait2}
    \end{subfigure}
    \caption{Performance of policies $\boldsymbol{\hat{\pi}}$ and $\boldsymbol{{\pi}^s}$: 2 flows with arrival processes $\mathcal{A}_1=(\eta,10/\eta,300/\eta)$ and $\mathcal{A}_2=(\eta,30/\eta,300/\eta)$; $\alpha_1 = 0.2$, $\alpha_2 = 0.6$, $V=1000$ and  $\zeta=1$. }\label{fig:bursty}
\end{figure*}

 \subsection{Comparison of Policies $\boldsymbol{\hat{\pi}}$ and $\boldsymbol{{\pi}^s}$ \label{sec:sim_cp1}}
  \begin{figure}[!t]
    \centering
    \includegraphics[width=0.4\linewidth]{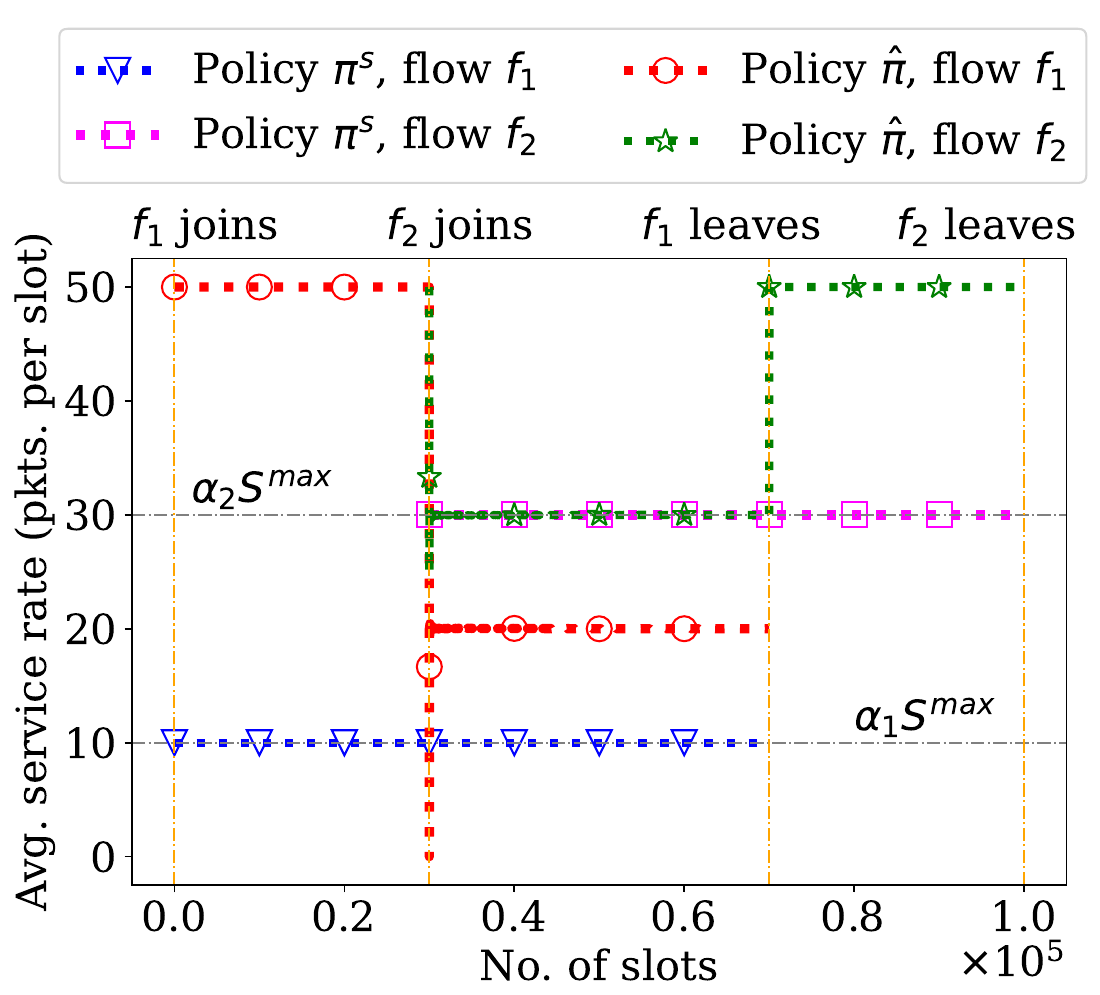}
    \caption{Average service rate under policies $\boldsymbol{\hat{\pi}}$ and $\boldsymbol{{\pi}^s}$; 2 flows with arrival process $\mathcal{A}=(1,20,300)$; $\alpha_1 = 0.2$, $\alpha_2 = 0.6$, $V=1000$ and $\zeta=1$.} \label{fig:df_service}
\end{figure}
We first consider the service requirements and arrival process as in Combination~3 of Table~\ref{tab:parameters}. Results of the simulation are presented in Fig.~\ref{fig:bursty}. When the traffic is not bursty, i.e., $\eta = 1$, weighted time-average drops are similar for both policies as the average packet arrival rate is less than the guaranteed average service rate (refer to Fig. \ref{fig:bursty_drop}). We observe that burstiness ($\eta = 10$) causes policy $\boldsymbol{{\pi}^s}$ to drop slightly more packets due to inflexibility in service rate allocation. Reduced service rates also lead to higher wait times at gNodeB under policy $\boldsymbol{{\pi}^s}$ (refer to Figs.~\ref{fig:bursty_wait1} and \ref{fig:bursty_wait2}).

\noindent
 \textbf{Remark:} \textit{Lack of flow isolation can lead to slightly larger delays for non-bursty flow when the average packet arrival rate is less than its average guaranteed packet service rate. As remarked in Sec.~\ref{sec:static}, this 
difference arises due to the dynamic rate allocation mechanism in policy $\boldsymbol{\hat{\pi}}$.}

 Next, we consider two flows $f_1$ and $f_2$ with packet arrival process $\mathcal{A}=(1,20,300)$. However, unlike before, these flows can enter and leave the system. Flow $f_1$ is the only flow in the system for the first $3 \times 10^4$ slots. In slot $3 \times 10^4$, flow $f_2$ joins the system. Subsequently, flows $f_1$ and $f_2$ leave in slots $7 \times 10^4$ and $10^5$, respectively. 
 Time-averaged\footnotemark{} service rates obtained by policies $\boldsymbol{\hat{\pi}}$ and $\boldsymbol{{\pi}^s}$ in this dynamic setting are presented in Fig.~\ref{fig:df_service}.
 
 \footnotetext{To better illustrate the adaptability of the system, averaging is reset when a flow joins/leaves the system. }

 When there is only one flow in the system, policy $\boldsymbol{\hat{\pi}}$ allocates the entire available capacity to it.  If there are multiple flows, policy $\boldsymbol{\hat{\pi}}$ ensures that each flow gets the minimum guaranteed average service rate. In addition to this, spare capacity is shared among co-existing flows based on their arrival process. On the other hand, for each flow, policy $\boldsymbol{{\pi}^s}$ allocates a fixed portion of the available capacity at all times. While such a scheme provides an average service rate guarantee independent of flows in the system, it comes at the expense of reduced system utilization.

\section{Impact of Closed-loop Flow Rate Control \label{sec:congestion}}

\begin{figure}[!t]
    \centering
    \includegraphics[width=0.6\linewidth]{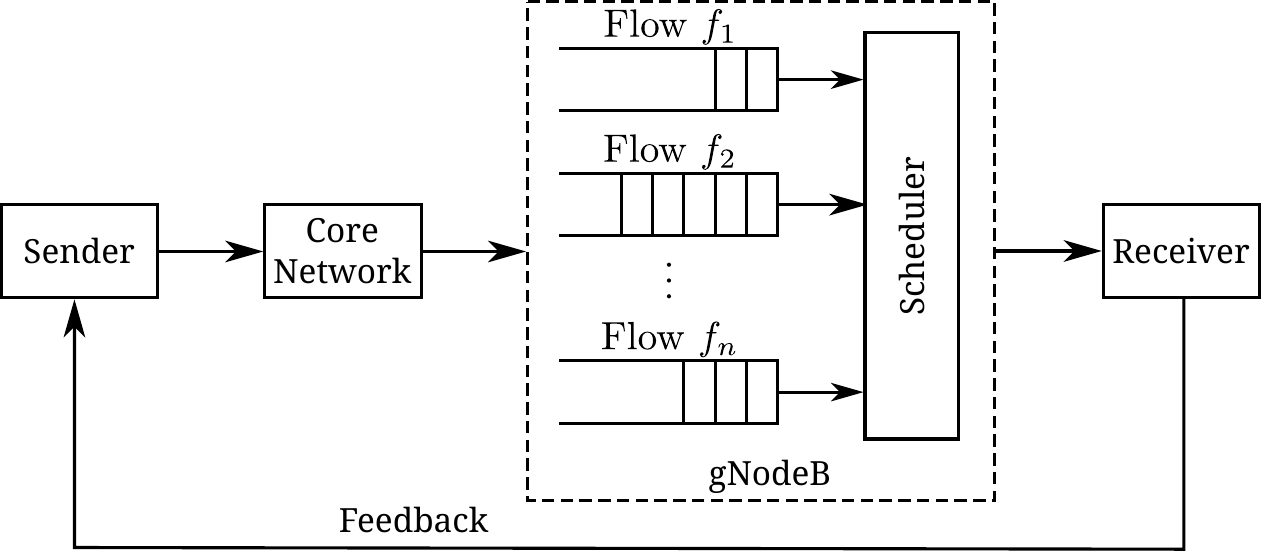}
    \caption{System with closed-loop flow rate control.} \label{fig:feedback}
\end{figure}

Injecting an unreasonably large number of packets into the network can lead to congestion which in turn can lead to large delays, packet drops, and bandwidth wastage. Consequently, QoS-aware flows often employ closed-loop mechanisms that achieve the best end-to-end flow rates
without overwhelming the underlying network. Therefore, in this section, we study the impact of a simple closed-loop flow rate controller on our policies $\boldsymbol{\hat{\pi}}$ and $\boldsymbol{\pi^s}$. A high-level representation of the system studied is presented in Fig.~\ref{fig:feedback}.

In \cite{goyal2017rethinking}, the authors note that a well-designed signaling scheme along with appropriate policy can perform better than \emph{Active Queue Management (AQM)}. Motivated by this, we consider a signaling mechanism in which the source is notified explicitly about each packet delivery and each drop decision. Packet delivery can be notified to the source via \emph{Acknowledgment (ACK)}. Whereas, a drop decision can be conveyed using a \emph{Negative Acknowledgement (NACK)} \cite{fox1989tcp,nack2005}. A NACK can be enabled via any of the following approaches.
\begin{enumerate}
    \item gNodeB uses \emph{Explicit Congestion Notification (ECN)} and mark ECN / feedback bit \cite{floyd1994tcp,goyal2017rethinking}. Upon receiving a marked packet, the receiver sends a NACK to the source.
    \item gNodeB can send a corrupted version of a packet to the receiver. Upon reception of such a packet, the receiver sends a NACK to the source \cite{almes1999one}. 
    \item gNodeB can spoof itself as the receiver and send a NACK to the source for each drop decision.
\end{enumerate}

In slot $t$, we assume that the source of a flow $f$ generates packets according to a Poisson distribution with mean $\lambda_f(t)$. Let $\mu^{+}_f(t)$ and $\mu^{-}_f(t)$ denote the number of ACKs and NACKs received by the source of flow $f$ in slot $t$. Then, for all $t \geq 0$, the mean of the Poisson distribution follows the following recursive equation
\begin{align}
\lambda_f(t+1)= \max \left\{ \frac{\lambda_f(t)+ 0.05 \times \mu^{+}_f(t)}{2^{\mu^{-}_f(t)}},1 \right\}   \label{eq:congetion_eq}
\end{align}
with $\lambda_f(0) = 1$. Eq~\eqref{eq:congetion_eq} is based on the \emph{Additive Increase Multiplicative Decrease (AIMD)} algorithm --- a feedback control algorithm best known for its use in TCP congestion control. 

\begin{figure}[!t]
    \begin{subfigure}{\linewidth}
    \centering
    \includegraphics[width=0.4\linewidth]{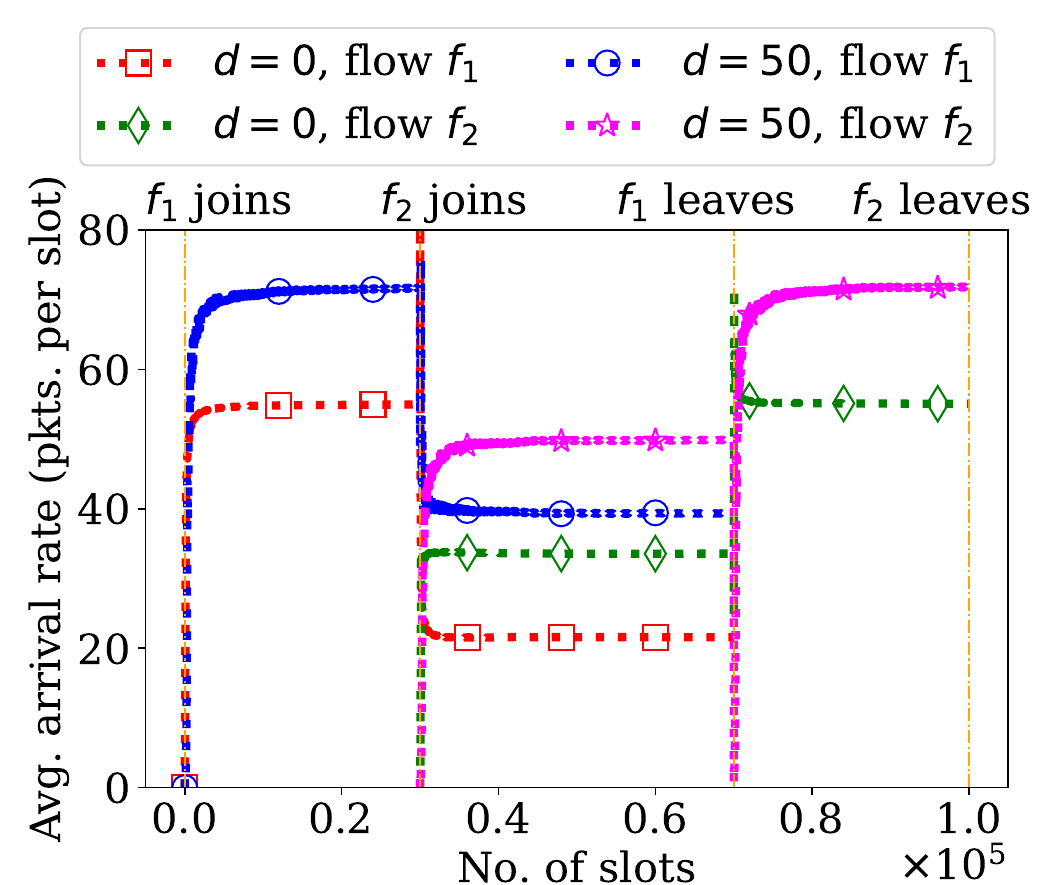}
    \caption{Average packet arrival rate} \label{fig:feedback_avg_arr}
    \end{subfigure}
    \begin{subfigure}{\linewidth}
    \centering
    \includegraphics[width=0.4\linewidth]{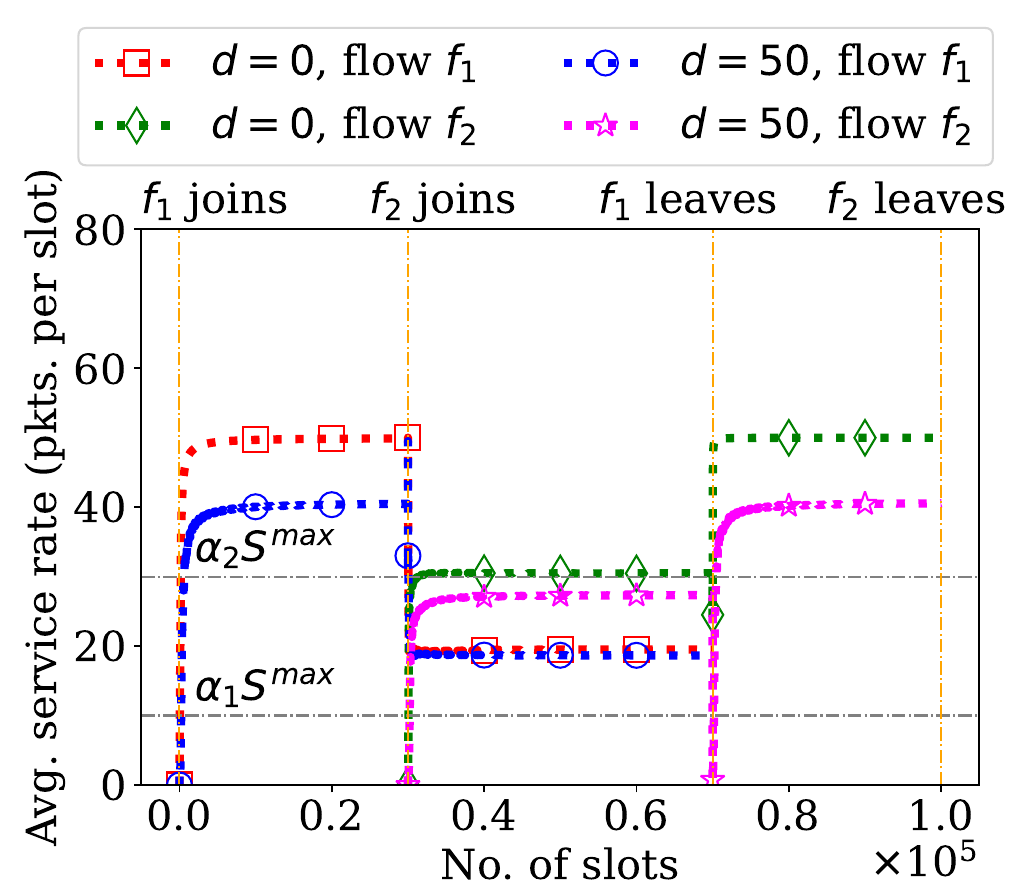}
    \caption{Average number of packets served} \label{fig:feedback_avg_ser}
    \end{subfigure}
     \caption{Time-averaged rates under policy $\boldsymbol{\hat{\pi}}$ with a feedback delay of $d$ slots: $\alpha_1 = 0.2$, $\alpha_2 = 0.6$, $V=1000$ and $\zeta=1$.} \label{fig:feedback_plot}
\end{figure}

\begin{figure}[!t]
    \begin{subfigure}{\linewidth}
    \centering
    \includegraphics[width=0.4\linewidth]{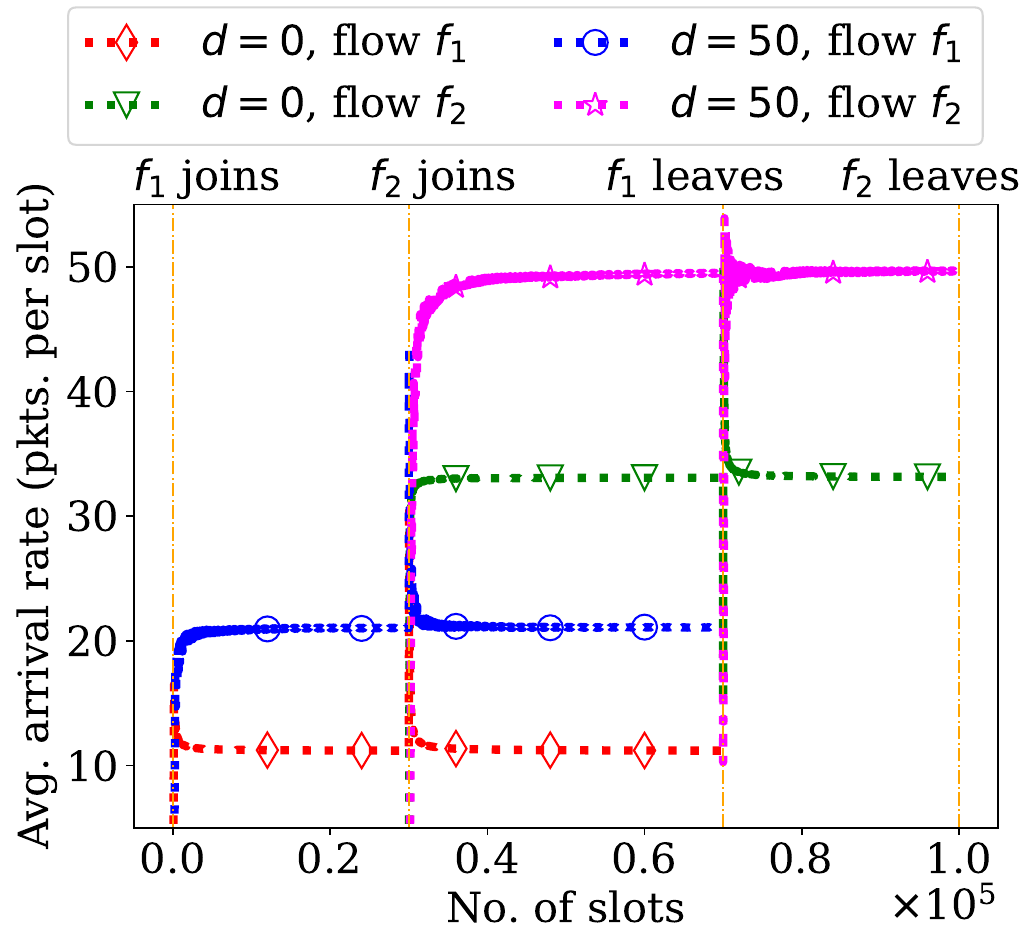}
    \caption{Average packet arrival rate} \label{fig:feedback_pis_avg_arr}
    \end{subfigure}
    \begin{subfigure}{\linewidth}
    \centering
    \includegraphics[width=0.4\linewidth]{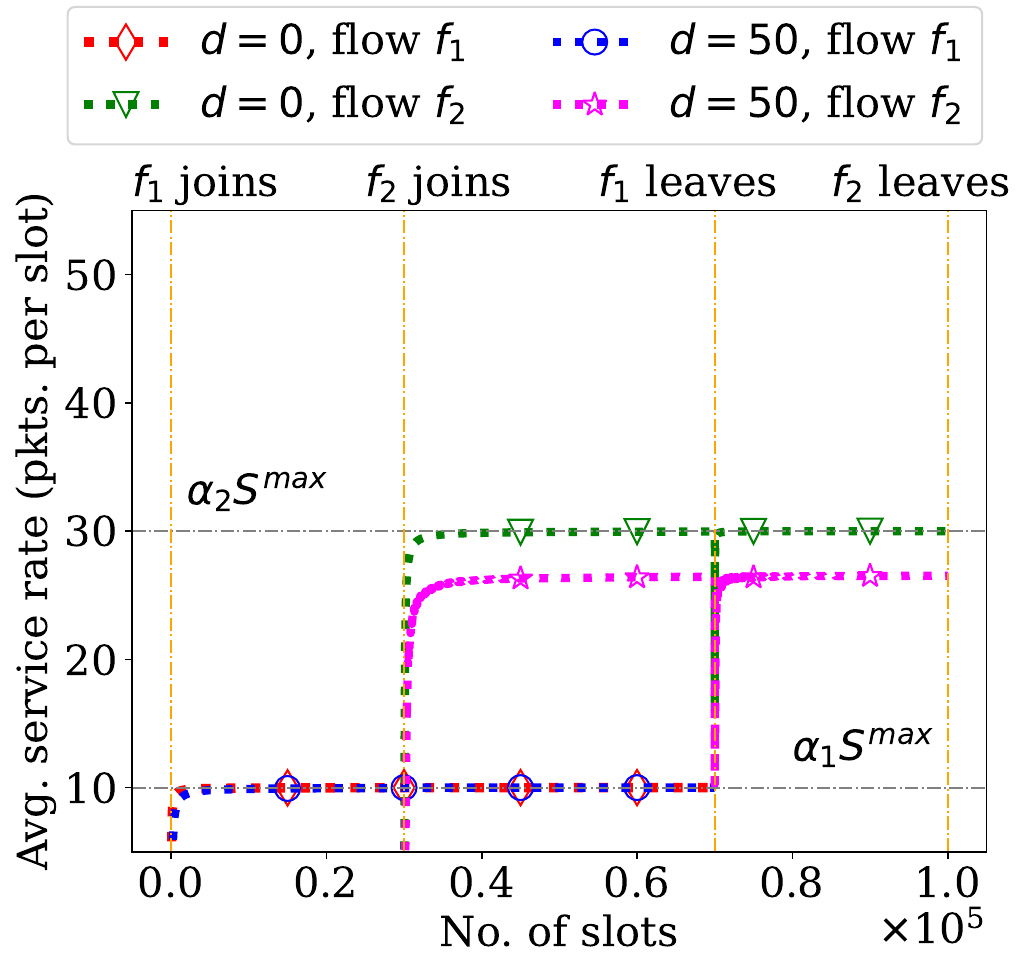}
    \caption{Average number of packet served} \label{fig:feedback_pis_avg_ser}
    \end{subfigure}
     \caption{Time-averaged rates under policy $\boldsymbol{{\pi}^{s}}$ with a feedback delay of $d$ slots: $\alpha_1  = 0.2$, $\alpha_2 = 0.6$, $V=1000$ and $\zeta=1$.} \label{fig:feedback_pis}
\end{figure}

To study the interplay between the closed-loop flow rate control and QoS-aware scheduling with feedback delays\footnote{We consider a feedback delay of $d$ slot. This delay may be arbitrarily distributed among the various components of the closed-loop system in Fig.~\ref{fig:feedback}.}, we consider two flows $f_1$ and $f_2$. Flow $f_1$ is the only flow in the system for the first $3 \times 10^4$ slots. In slot $3 \times 10^4$, flow $f_2$ joins the system. Subsequently, flows $f_1$ and $f_2$ leave in slots $7 \times 10^4$ and $10^5$, respectively. The service requirements of these flows are chosen as $\alpha_1 = 0.2$ and $\alpha_2 = 0.6$, respectively. The time-averaged\footnotemark[2] arrival and admission rate of these flows under  policies $\boldsymbol{\hat{\pi}}$ and $\boldsymbol{{\pi}^s}$ are presented in Figs.~\ref{fig:feedback_plot} and \ref{fig:feedback_pis}, respectively.

\begin{figure*}[!t]
\centering
    \begin{subfigure}{0.4\linewidth}
    \centering
    \includegraphics[width=\linewidth]{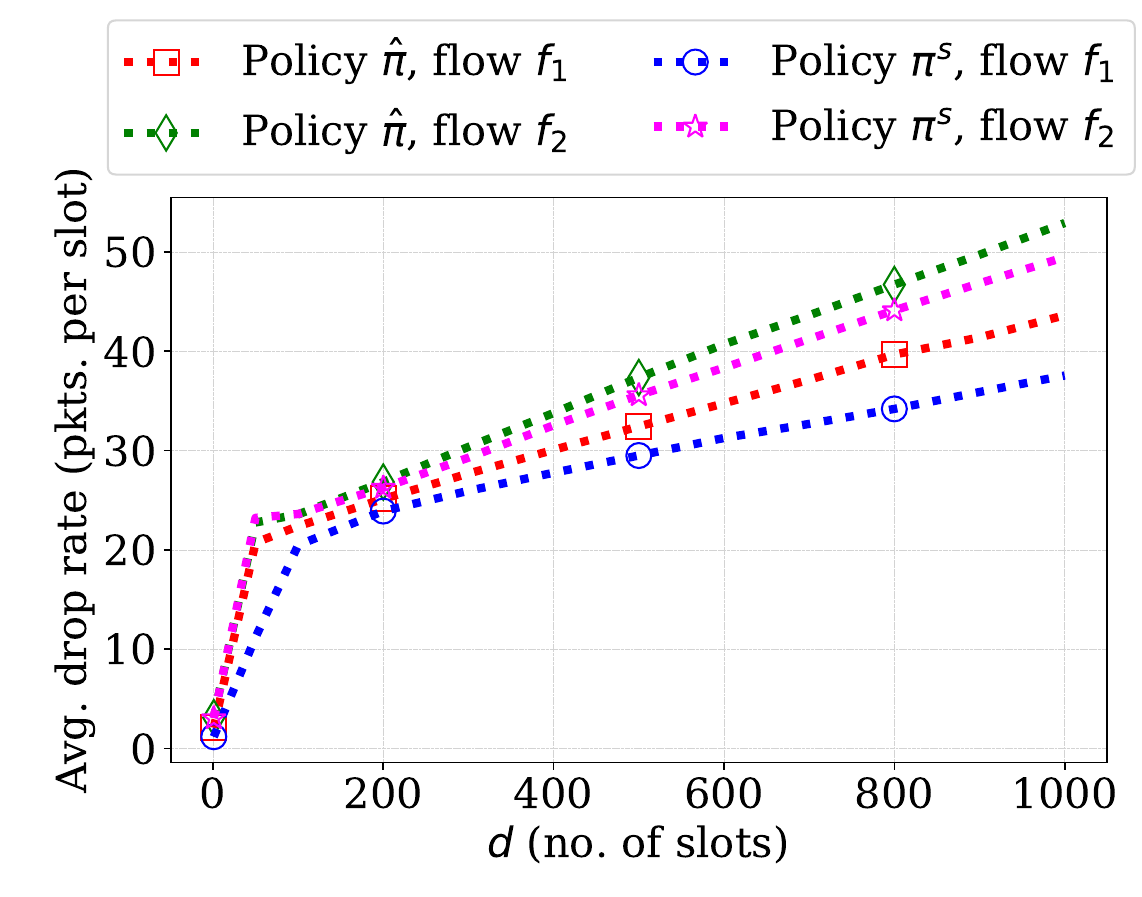}
    \caption{Average packet drop rate }\label{fig:feedback_d_drops}
    \end{subfigure}
    \begin{subfigure}{0.4\linewidth}
    \centering
    \includegraphics[width=\linewidth]{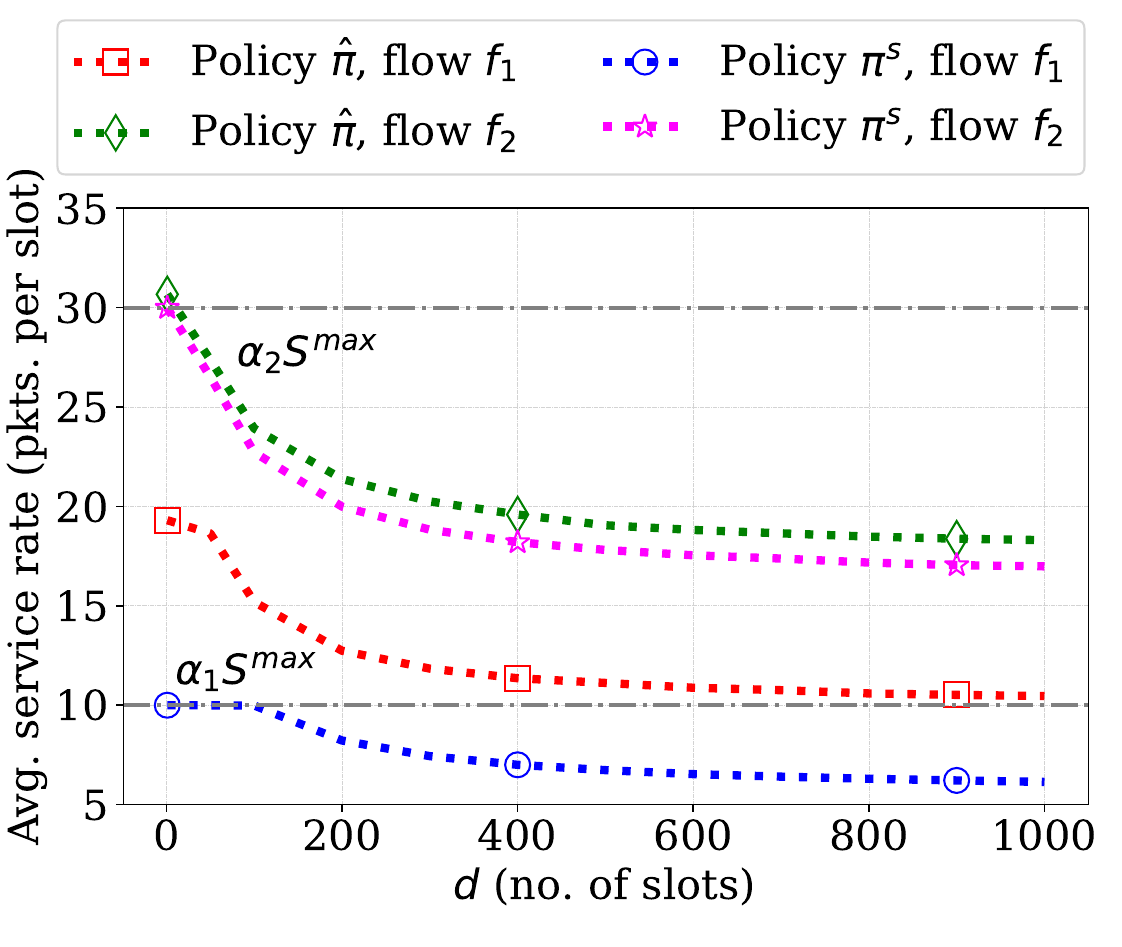}
    \caption{Average number of packets served} \label{fig:feedback_d_service}
    \end{subfigure}
       \begin{subfigure}{0.41\linewidth}
    \centering
    \includegraphics[width=\linewidth]{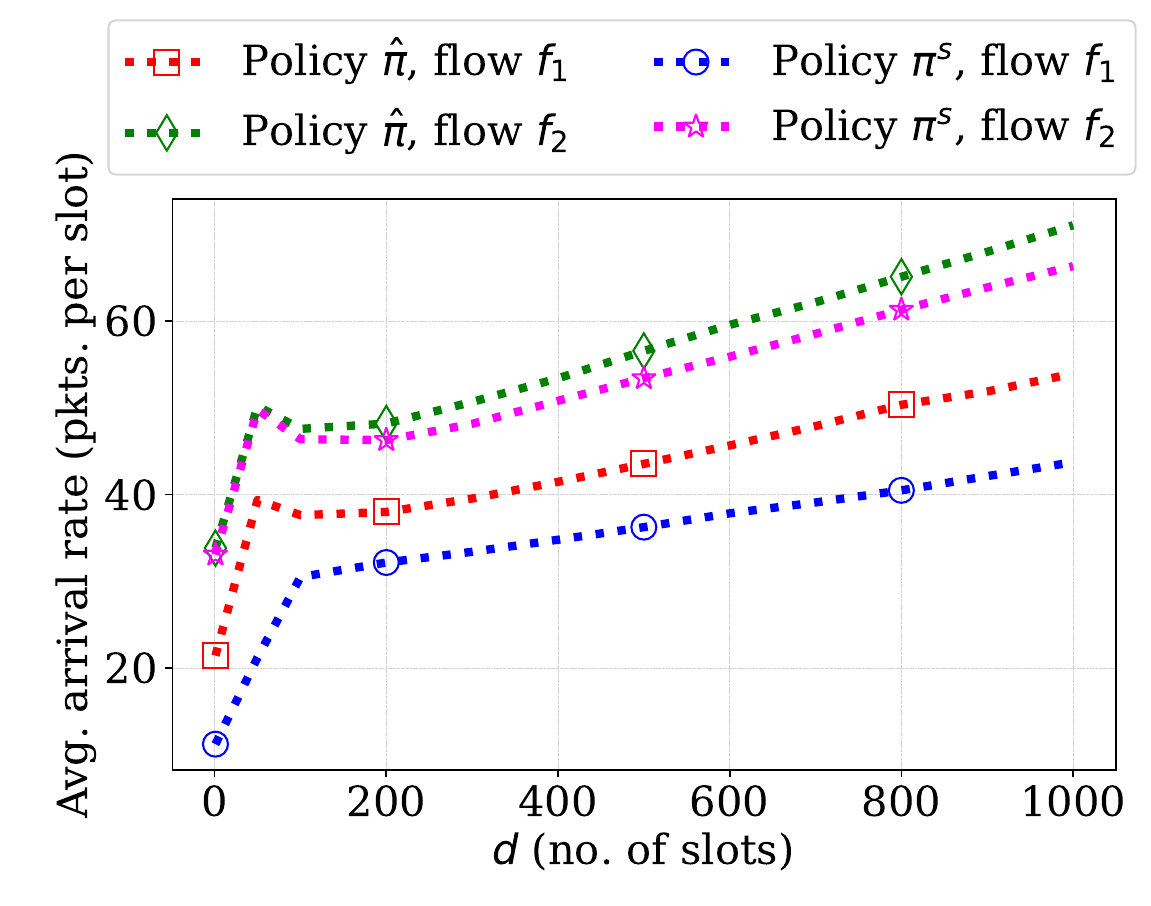}
    \caption{Average arrival rate} \label{fig:feedback_d_avg_arr}
    \end{subfigure}
     \caption{Average rates achieved with a feedback delay of $d$ slots: $\alpha_1  = 0.2$, $\alpha_2  = 0.6$, $V=1000$ and $\zeta=1$.}\label{fig:feedback_d}
\end{figure*}

\begin{figure*}[!t]
    \begin{subfigure}{0.32\linewidth}
    \centering
    \includegraphics[width=\linewidth]{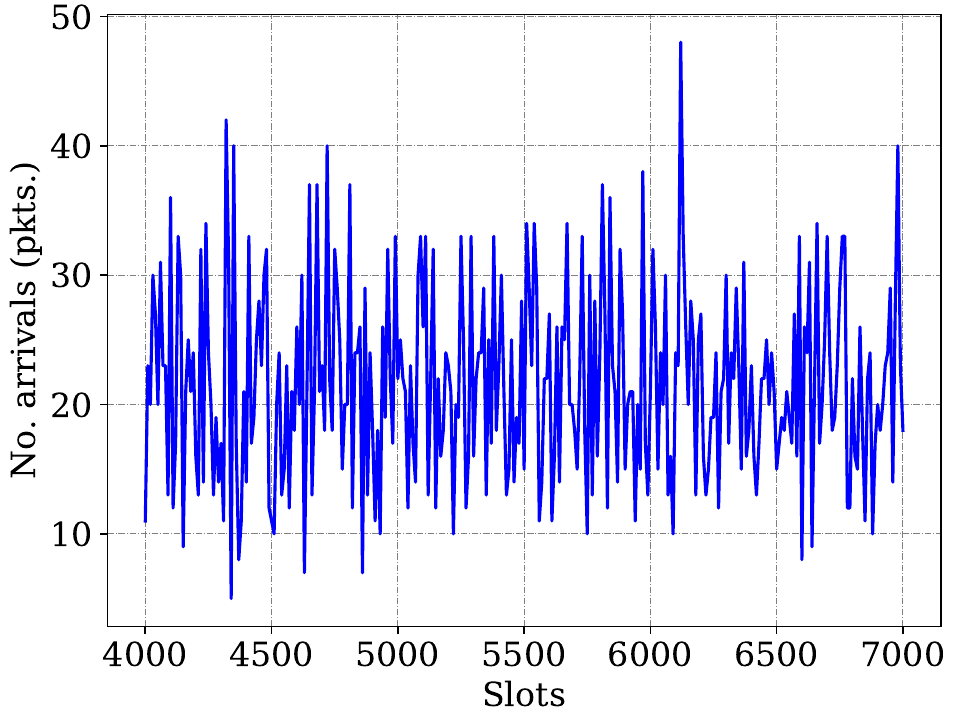}
    \caption{Number of packet arrivals: $d=0$.}\label{fig:feedback_d_0_arr}
    \end{subfigure}
    \begin{subfigure}{0.32\linewidth}
    \centering
    \includegraphics[width=\linewidth]{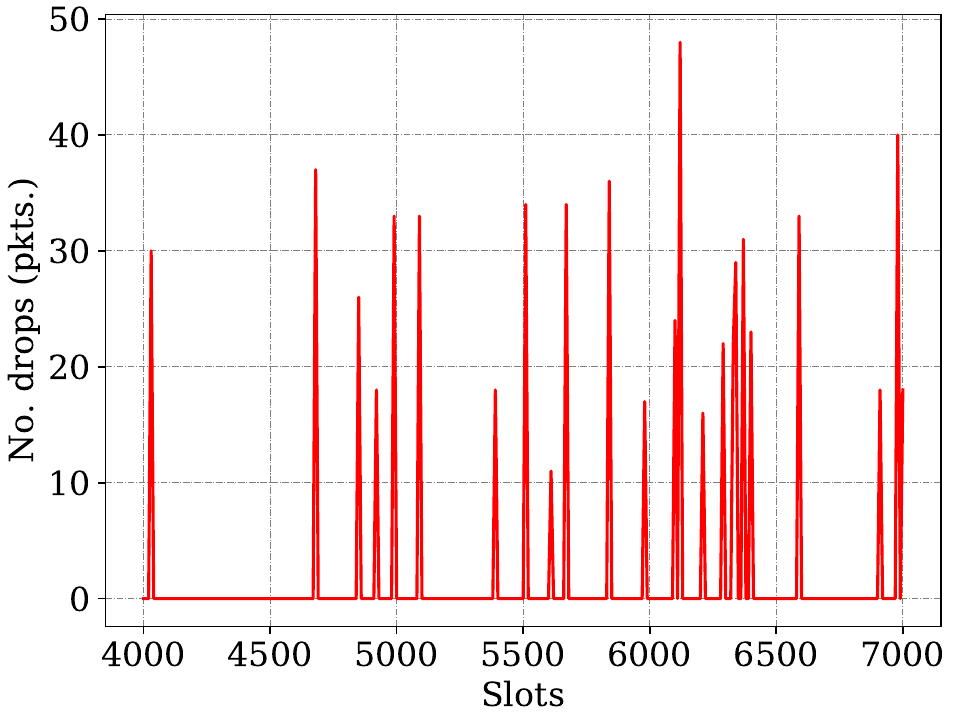}
    \caption{Number of packet drops: $d=0$.} \label{fig:feedback_d_0_drop}
    \end{subfigure}
    \begin{subfigure}{0.32\linewidth}
    \centering
    \includegraphics[width=\linewidth]{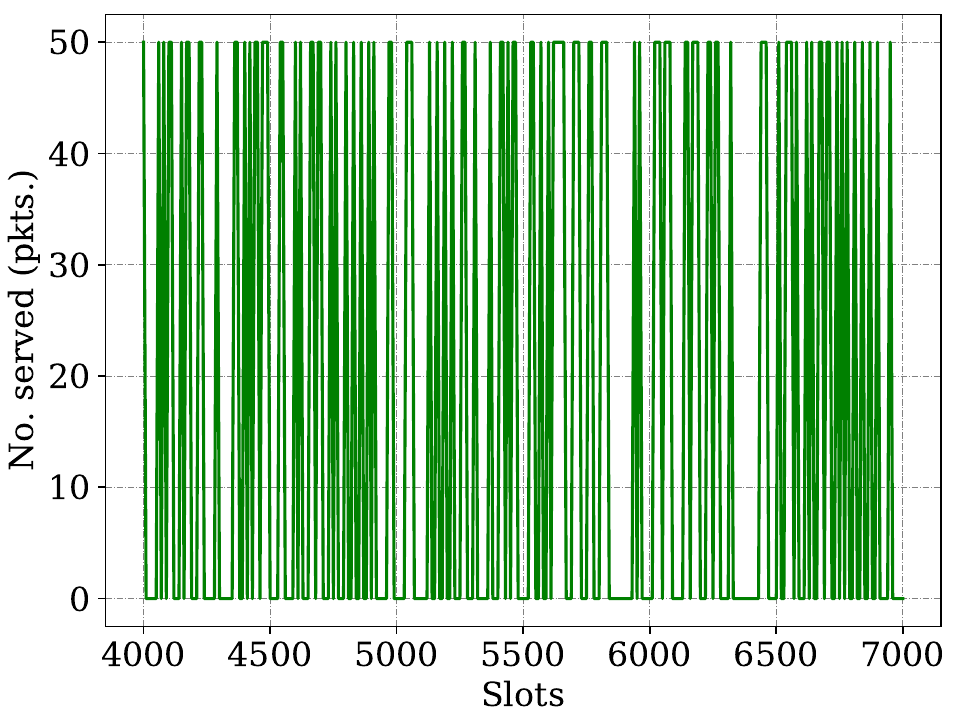}
    \caption{Number of packets served: $d=0$.}\label{fig:feedback_d_0_ser}
    \end{subfigure}
    \vspace{1mm}
    
    \begin{subfigure}{0.32\linewidth}
    \centering
    \includegraphics[width=\linewidth]{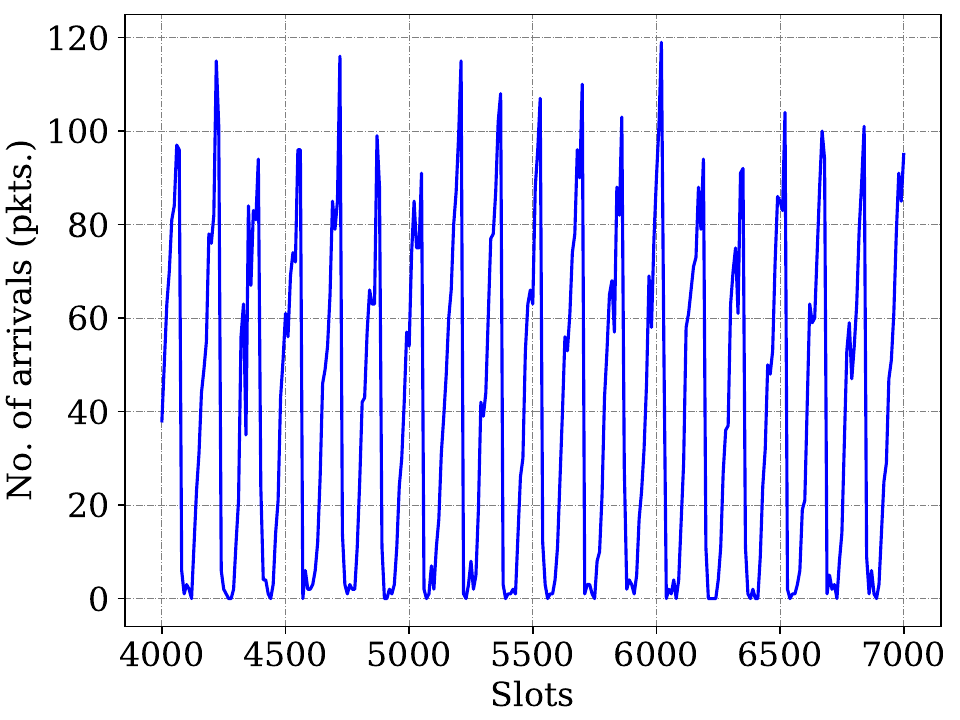}
    \caption{Number of packet arrivals: $d=50$.}\label{fig:feedback_d_50_arr}
    \end{subfigure}
    \begin{subfigure}{0.32\linewidth}
    \centering
    \includegraphics[width=\linewidth]{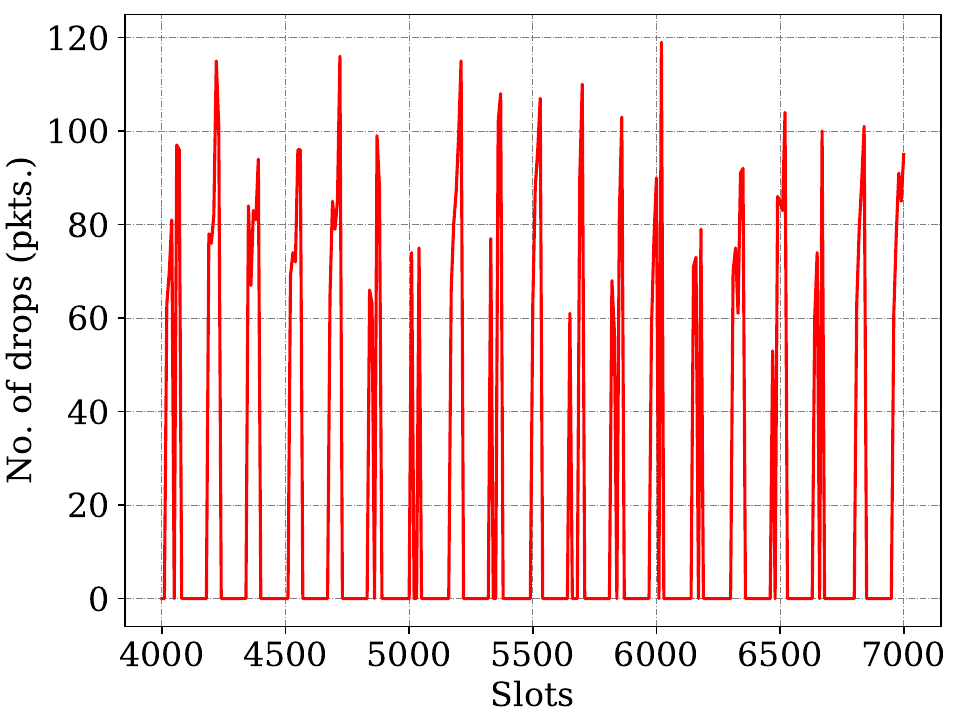}
    \caption{Number of packets dropped: $d=50$. }\label{fig:feedback_d_50_drop}
    \end{subfigure}
    \begin{subfigure}{0.32\linewidth}
    \centering
    \includegraphics[width=\linewidth]{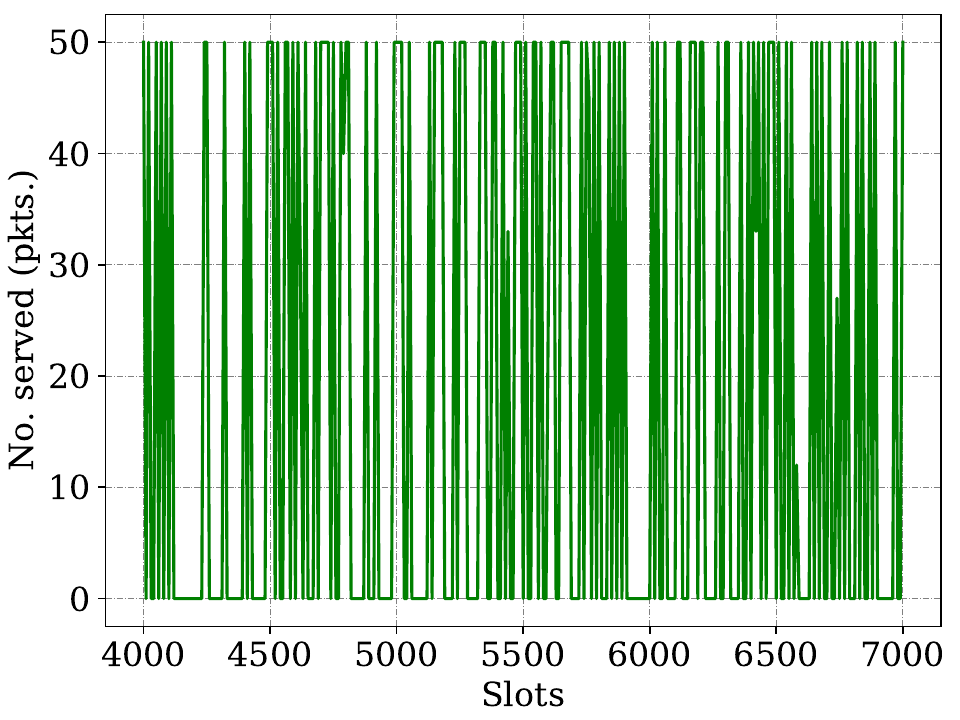}
    \caption{Number of packets served: $d=50$.}\label{fig:feedback_d_50_ser}
    \end{subfigure}
    \vspace{1pt}
    
    \begin{subfigure}{0.32\linewidth}
    \centering
    \includegraphics[width=\linewidth]{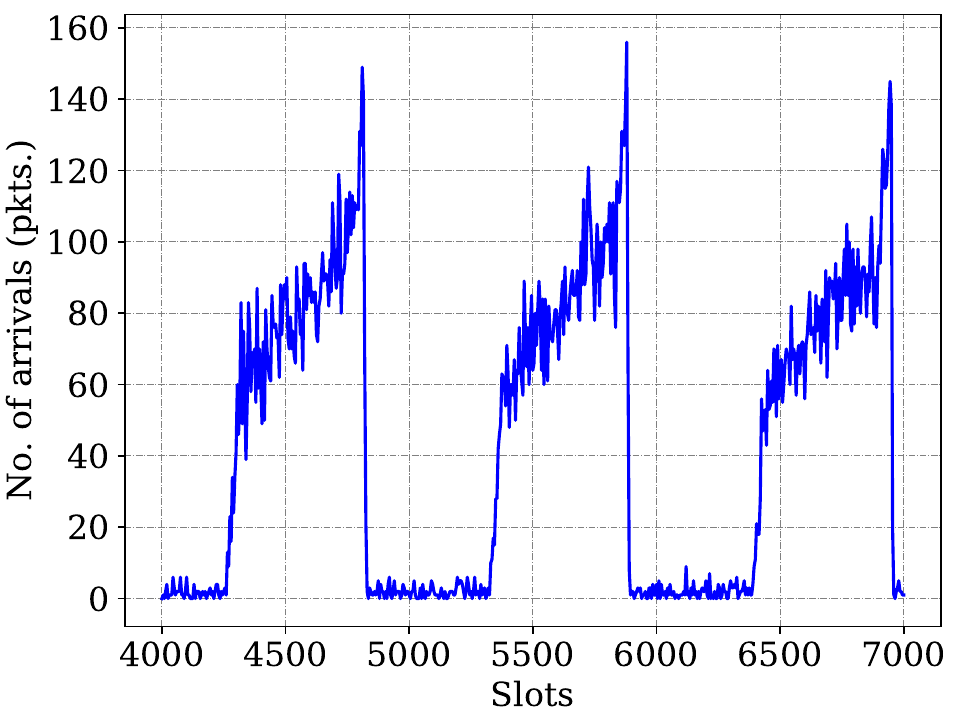}
    \caption{Number of packet arrivals: $d=500$.}\label{fig:feedback_d_500_arr}
    \end{subfigure}
    \begin{subfigure}{0.32\linewidth}
    \centering
    \includegraphics[width=\linewidth]{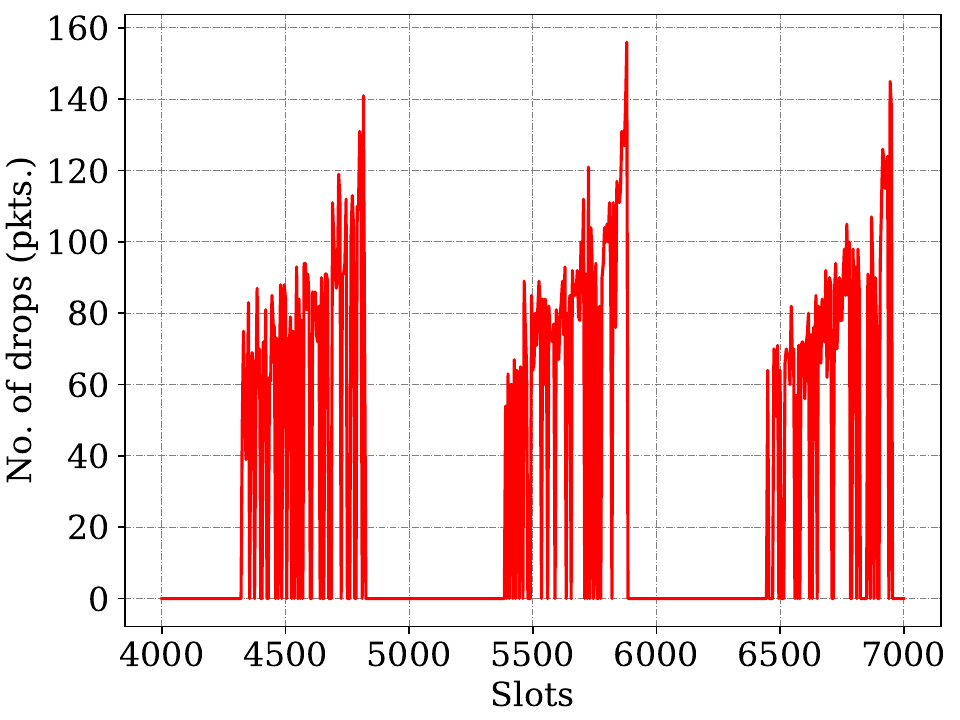}
    \caption{Number of packet drops: $d=500$.}\label{fig:feedback_d_500_drop}
    \end{subfigure}
    \begin{subfigure}{0.32\linewidth}
    \centering
    \includegraphics[width=\linewidth]{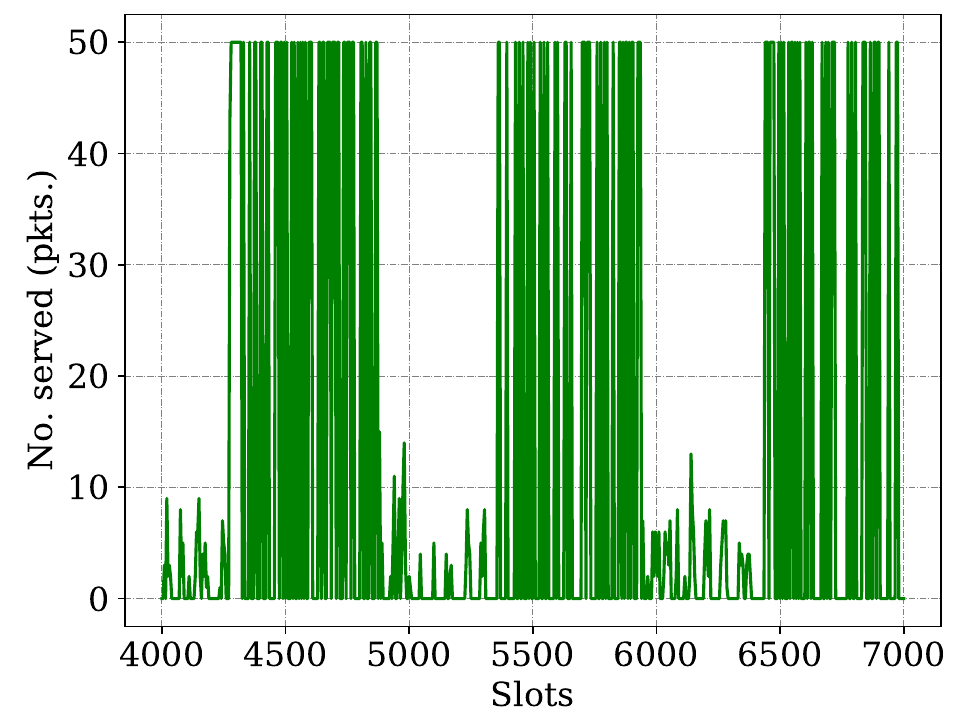}
    \caption{Number of packets served: $d=500$. }\label{fig:feedback_d_500_ser}
    \end{subfigure}
     \caption{Snapshot of flow $f_1$'s processes: policy $\boldsymbol{\hat{\pi}}$, feedback delay of $d$ slots, $\alpha_1  = 0.2$, $\alpha_2  = 0.6$, $V=1000$ and $\zeta=1$.}\label{fig:feedback_d_detail}
\end{figure*}

When the feedback reaches the source without any delays i.e., $d=0$, policy $\boldsymbol{\hat{\pi}}$ can achieve an average end-to-end flow rate equal to the average system capacity in a single flow system (refer to Fig.~\ref{fig:feedback_avg_ser}). On the other hand, policy $\boldsymbol{{\pi}^s}$, only achieves an end-to-end flow rate equal to the minimum guaranteed service rate (refer to Fig.~\ref{fig:feedback_pis_avg_ser}). 
Feedback with zero delays also allows policies $\boldsymbol{\hat{\pi}}$ and $\boldsymbol{{\pi}^s}$ to share the system capacity among co-existing flows as discussed in Sec.~\ref{sec:sim_cp1}.
Under both policies, flows' packet arrival rate (refer to Fig.~\ref{fig:feedback_avg_arr} and \ref{fig:feedback_pis_avg_arr}) exceeds its service rate, leading to a drop of at least $10 \, pkts/slot$ at the gNodeB. This is an inevitable artifact of the loose coupling between the closed-loop flow rate control and QoS-aware scheduling.

A feedback delay of just $d=50$ slots causes both policies to achieve lower service and higher arrival rates, compounding in larger packet drop rates compared to the scenario without feedback delays. To better understand the cause of this issue, we simulate two co-existing flows $f_1$ and $f_2$ over a large range of feedback delays. The results of this simulation are presented in Fig.~\ref{fig:feedback_d}. We observe from Figs.\ref{fig:feedback_d_drops} and \ref{fig:feedback_d_service} that, irrespective of flow isolation, average drop, and arrival rates are increasing piece-wise linear functions of the feedback delay. Slopes of these curves exhibit a transition at $d=50$. Based on a preliminary investigation, we believe that the transition point is determined by the value of parameter $V$.  In the future, we plan to take up a detailed study to understand and possibly quantify this dependence.

 In Fig.\ref{fig:feedback_d_detail}, we present a snapshot of flow $f_1$'s packet arrival, drop, and service processes under policy $\boldsymbol{\hat{\pi}}$ for feedback delays $d=0$, $d=50$ and $d=500$. Without feedback delays, the arrival process oscillates between 10 and 40 packets (refer to Fig.~\ref{fig:feedback_d_0_arr}). With a delay of $50$ slots, the frequency of oscillation decreases, the range of oscillations almost quadruples, and we periodically see slots with no packet arrivals (refer to Fig.~\ref{fig:feedback_d_50_arr}). When the delay increases to $500$ slots, the packet arrivals process exhibits a periodic behavior with period $2d = 1000$ slots (refer to Fig.~\ref{fig:feedback_d_500_arr}). For the first half of each period, just a few packets arrive in each slot. This happens due to the reception of delayed \emph{NACK}s from the immediately previous period. On the other hand, the second half of each period sees a linear increase from 60 to 140 packets. Since the maximum service rate at the gNodeB is $50 \, pkts/slot$, such a large inflow of packets causes frequent drops (refer to Fig.~\ref{fig:feedback_d_500_drop}), in turn resulting in a larger number of NACKs in the first half of the subsequent period. Packet drop decisions are sporadic without feedback delays, but become frequent and closely packed as the delay increases (refer to Figs.~\ref{fig:feedback_d_0_drop}, \ref{fig:feedback_d_50_drop} and \ref{fig:feedback_d_500_drop}).

The packet arrival process has a periodic HIGH-LOW pattern in settings with large feedback delays. Further, gNodeB just has a few packets to serve in the LOW periods of the arrival process (refer to Fig.~\ref{fig:feedback_d_500_ser}). Due to these reasons, despite an increase in the number of packet arrivals, the average number of packets served decreases with an increase in the feedback delay (refer to Fig.~\ref{fig:feedback_d_service}). We would like to note that this issue persists even if the flows are isolated.

\section{Conclusion and Future Work \label{sec:conclude}}

Servicing flows with different QoS requirements using a limited set of resources is a key challenge in RAN slicing. In this paper, we formulated a resource allocation problem to minimize a weighted long-term time average of packet drop decisions subject to average guaranteed service rate and queue stability constraints. We then presented two policies: the first one is a delay-guaranteed near-optimal admission and scheduling policy whose performance can be controlled with a couple of parameters, and the second one is a delay-guaranteed sub-optimal policy that provides flow isolation. 

We carried out extensive simulations to study the performance of our policies. We first studied the influence of systems and policy parameters. We then extended our study to a more realistic scenario where flows can enter and leave the system. In Section~\ref{sec:congestion}, we replicate these investigations to assess performance in the presence of a closed-loop flow rate. While we were unable to find any pertinent real-world data sets for validating our policies, we believe our simulation outcomes offer a qualitative glimpse into the efficacy of our proposed strategies in real-world 5G networks.

In future work, we plan to investigate the impact of other well-known feedback mechanisms on our QoS-aware scheduling policies. We also plan to explore QoS-aware scheduling policies that achieve minimal packet drops and better service rates in networks with large feedback delays. 
\appendices

\setcounter{equation}{0}
\renewcommand\theequation{A. \arabic{equation}}

\section{}

\subsection{Proof of Proposition~\ref{prop:serv_const} \label{sec:serv_const_proof}}
\noindent
From Eq.~\eqref{eq:yq}, we have $Y_i(t+1) - Y_i(t) \geq \alpha_i S(t) -S_i(t)$. Summing both sides of this equation from 1 to $T$, rearranging the terms, dividing throughout by $T$, and taking the limit superior on both sides, we have
$
- \liminf_{T \to \infty} \frac{1}{T} \sum^T_{t=1} [S_i(t) - \alpha_i S(t)] = \limsup_{T \to \infty} \frac{1}{T} \sum^T_{t=1} [\alpha_i S(t) -S_i(t)] \leq \limsup_{T \to \infty}  \frac{Y_i(T+1)}{T}
$. If queue $\mathcal{Y}_i$ is rate stable, we have $\limsup_{T \to \infty}  \frac{Y_i(T+1)}{T}=0$. \hfill \IEEEQED

\subsection{Derivation of Inequality~\ref{eq:drift_p} \label{sec:drift}}
\noindent
From Eq.~\eqref{eq:dq}, we have
\begin{align}
&Q_i(t+1)^2-Q_i(t)^2 = ([Q_i(t)-S_i(t)-D_i(t)]^++A_i(t))^2-Q_i(t)^2 \nonumber \\
&\leq A_i(t)^2+ D_i(t)^2+S_i(t)^2 -2Q_i(t)[S_i(t)+D_i(t)] \nonumber \\ 
& \hspace{5mm} +2A_i(t)[Q_i(t)-S_i(t)-D_i(t)]^+ \nonumber \\
& \overset{a}{\leq} (A_i^{max} + S^{max} + D_i^{max})^2  \nonumber \\
& \hspace{5mm} + 2Q_i(t)[A_i(t)-S_i(t)-D_i(t)] \label{eq:lap_eq_part1}
\end{align}

where (a) follows because  $[x^+]^2\leq x^2\hspace{1mm} \forall x \in \mathbb{R}$ and $[x-y]^+\leq x \hspace{1mm} \forall x,y \in \mathbb{R}^+$. Similarly, from Eq.~\eqref{eq:yq}, we have
\begin{align}
Y_i(t+1)^2-Y_i(t)^2 \leq 2 (S^{max})^2 + 2Y_i(t)[\alpha_i S(t)-S_i(t)] \label{eq:lap_eq_part2}
\end{align}

\noindent
Finally, from Eq.~\eqref{eq:zq}, we have
\begin{align}
&Z_i(t+1)^2-Z_i(t)^2 \leq   2\zeta^2 (S^{max}+D^{max}_i)^2 \nonumber\\
& \hspace{10mm} +2\zeta Z_i(t)[\alpha_i S(t) \mathbb{I}_i(t) - S_i(t) - D_i(t)] \label{eq:lap_eq_part3}
\end{align}

Substituting Inequalities~\eqref{eq:lap_eq_part1}, \eqref{eq:lap_eq_part2} and \eqref{eq:lap_eq_part3} in the expression for $L(t + 1) - L(t)$, we obtain the desired inequality. 

\hfill \IEEEQED

\subsection{Proof of Proposition~\ref{prop:pio_feasible}
\label{sec:pio_feasible_proof}} 

Let $\{\overline{Q}_i(t), i \in \mathcal{N} \}^{\infty}_{t=1}$, $\{\overline{Y}_i(t), i \in \mathcal{N} \}^{\infty}_{t=1}$, and $\{\overline{Z}_i(t), i \in \mathcal{N} \}^{\infty}_{t=1}$ denote the length of data, virtual and persistent queues under policy $\boldsymbol{\overline{\pi}}$. The following lemma presents an upper bound on the length of these queues.

\begin{lem} \label{lem:data_queue_bound}
For all $ i \in \mathcal{N}, t \geq 1$, we have $\overline{Q}_i(t)\leq Vw_i+A_i^{max}$.
\end{lem}
\begin{IEEEproof}
When $\overline{Q}_i(t)\leq Vw_i$, from \eqref{eq:dq}, we have
\begin{align*}
\overline{Q}_i(t+1)=&[\overline{Q}_i(t)-\overline{S}_i(t)-\overline{D}_i(t)]^+ +A_i(t)\\
\leq& \overline{Q}_i(t)+A_i(t) \leq Vw_i+A_i^{max}
 \end{align*}
When $\overline{Q}_i(t)> Vw_i$, due to the drop rule \eqref{eq:p2_dropdecision}, we have 

\noindent
\textit{Case I}: If $\overline{Q}_i(t) < A^{max}_i$, then 
\begin{align*}
\overline{Q}_i(t+1) &\leq [\overline{Q}_i(t) -D_i^{max}]^++A_i(t) \\
&\leq [A^{max}_i -D_i^{max}]^++A_i(t) \overset{a}{\leq} A^{max}_i
 \end{align*}
\noindent
where (a) follows because $D_i^{max}\geq A^{max}_i$.

\noindent
\textit{Case II}: If $\overline{Q}_i(t) = A^{max}_i + \epsilon$ where $\epsilon \geq 0$, then
\begin{align*}
\overline{Q}_i(t+1) &\leq [A^{max}_i + \epsilon -D_i^{max}]^++A_i(t) \\
&\overset{(b)}{\leq} \epsilon + A^{max}_i \leq \overline{Q}_i(t)
 \end{align*}
 where (b) follows because $D_i^{max}\geq A^{max}_i$.
 
 Since $\overline{Q}_i(0) = 0$, the above arguments can be recursively applied to prove that flow $i$'s data queue length is bounded above by $V w_i + A^{max}_i$.
\end{IEEEproof}  

\begin{lem} \label{lem:z_queue_bound}
For all $i \in \mathcal{N}, t \geq 1$, $\overline{Z}_i(t)\leq  {Vw_i}/{\zeta}+\zeta \alpha_i S^{max}$.
\end{lem}
\begin{IEEEproof}
When $\zeta \overline{Z}_i(t)\leq Vw_i$, from \eqref{eq:zq}, we have
\begin{align*}
\overline{Z}_i(t+1) &= [ \overline{Z}_i(t)+\zeta \alpha_i S^{max} (t)\mathbb{I}_i(t)-\zeta \overline{S}_i(t)-\zeta \overline{D}_i(t)]^+\\
&\leq  \overline{Z}_i(t)+\zeta \alpha_i S(t) \mathbb{I}_i(t) \leq {Vw_i}/{\zeta}+\zeta \alpha_i S(t)\\
&\leq {Vw_i}/{\zeta}+\zeta \alpha_i S^{max}
 \end{align*}

On the other hand when $\zeta \overline{Z}_i(t) > Vw_i$, due to drop rule \eqref{eq:p2_dropdecision}, we have 
\begin{align*}
\overline{Z}_i(t+1)&=[ \overline{Z}_i(t)+\zeta \alpha_i S(t) \mathbb{I}_i(t)-\zeta \overline{S}_i(t)-\zeta D_i^{max}]^+\\
&\leq [ \overline{Z}_i(t)+\zeta \alpha_i S(t) -\zeta D_i^{max}]^+  \overset{b}{\leq} \overline{Z}_i(t)
 \end{align*}
where (b) follows because $D_i^{max}\geq\alpha_iS(t)$.
\end{IEEEproof}

\begin{lem} \label{lem:dataz_queue_bound}
For all $i \in \mathcal{N}, t \geq 1$, $\zeta \overline{Z}_i(t) + \overline{Q}_i(t) \leq  {V}w_i + \zeta^2 \alpha_i S^{max} +  A^{max}_i$.
\end{lem}
\begin{IEEEproof}
Along similar lines as proof of Lemmas~\ref{lem:data_queue_bound} and \ref{lem:z_queue_bound}.
\end{IEEEproof}

\begin{lem} \label{lem:y_queue_bound}
For all $i \in \mathcal{N}, t \geq 1$, $\overline{Y}_i(t)\leq  n(V+ \zeta^2 S^{max} + \max_{1 \leq i \leq n} A_i^{max}) + (n+1)S^{max}$.
\end{lem}
\begin{IEEEproof}
Let $\overline{i}(t) = \arg \max_{k \in \mathcal{N} } \zeta \overline{Z}_k(t)+\overline{Q}_k(t)+\overline{Y}_k(t)$ (tie can be broken using any arbitrary rule). Then, we have
$$\overline{Y}_i(t+1) = \begin{cases} [ \overline{Y}_i(t)+\alpha_i S(t)-S(t) ]^{+} & \textrm{if } i = \overline{i}(t) \\ 
\overline{Y}_i(t)+\alpha_i S(t) & \textrm{otherwise} \end{cases}$$

\noindent
Let us define the following time instants
\begin{align*}
t_{1} &= \min\{t > 0: \overline{W}_i(t) >\beta+S^{max} \textrm{ for some } i \in \mathcal{N}\} \\
t_{2} &= \min\{t > t_1: \overline{W}_i(t) \leq \beta+S^{max} \textrm{ for all } i \in \mathcal{N}\}
\end{align*}
where $\overline{W}_i(t) = \zeta \overline{Z}_i(t) + \overline{Q}_i(t)+\overline{Y}_i(t)$, and $\beta = V+ \zeta^2 S^{max} + \max_{1 \leq i \leq n} A_i^{max}$.

If $t_1=\infty$, then $\overline{W}_i(t) \leq \beta+S^{max} \, \forall i \in \mathcal{N}, t \geq 1$. Consequently, $\overline{Y}_i(t) \leq \beta+S^{max}$. Now, let us consider the case when $t_1 < \infty$. Then, for all $t \in [t_1,t_2)$, we have
\begin{align*}
\overline{Y}_{\overline{i}(t)}(t) = \overline{W}_{\overline{i}(t)}(t) - \zeta \overline{Z}_{\overline{i}(t)}(t) - \overline{Q}_{\overline{i}(t)}(t) 
& \overset{(a)}{>} S^{max}
\end{align*}
where (a) follows because $\overline{W}_{\overline{i}(t)}(t) = \max_{1 \leq i \leq n} \overline{W}_i(t) > \beta + S^{max}$, and $\zeta \overline{Z}_{\overline{i}(t)}(t) + \overline{Q}_{\overline{i}(t)}(t) \leq \beta$ (refer to Lemma~\ref{lem:dataz_queue_bound}).

\noindent
Consequently, for all $t \in [t_1,t_2)$, we have
\begin{align*}
& \sum^{n}_{i=1} \overline{Y}_i(t+1) = \sum^{n}_{i=1} \overline{Y}_i(t) + S(t) ( \sum^{n}_{i=1} \alpha_i - 1) \overset{(a)}{\leq} \sum^{n}_{i=1} \overline{Y}_i(t) \\
&\overset{(b)}{\leq} \sum^{n}_{i=1} \overline{Y}_i(t_1) 
= \hspace{-1mm} \sum^{n}_{i=1} [\overline{Y}_i(t_1-1) + \alpha_i S(t_1-1) - \overline{S}_i(t_1-1) ] \\
&\leq \sum^{n}_{i=1} \overline{Y}_i(t_1-1) + \alpha_i S(t_1-1) \overset{(c)}{\leq} n(\beta + S^{max})  + S^{max}
\end{align*}
where (a) holds because $\sum^{n}_{i=1} \alpha_i \leq 1$; (b) holds from a recursive application of inequality $\sum^{n}_{i=1} \overline{Y}_i(t+1) \leq \sum^{n}_{i=1} \overline{Y}_i(t) \, \forall t \in [t_1,t_2)$ and (c) holds because $\overline{Y}_i(t_1-1) \leq \beta + S^{max} \, \forall i \in \mathcal{N}$.

\begin{figure*}[!t]
\begin{align}
& \Delta \hat{L}(t) + V \hat{D}(t) \leq  C_1 + \zeta^2 C_2 + \sum^n_{i=1} \hat{D}_i(t) [V w_i - \zeta \hat{Z}_i(t) - \hat{Q}_i(t)] - \sum^n_{i=1} \overline{S}_i(t)[\zeta \hat{Z}_i(t) + \hat{Q}_i(t) + \hat{Y}_i(t)] \nonumber \\
& \sum^n_{i=1} + [\zeta \alpha_i  S(t) \hat{Z}_i(t) +  \hat{Q}_i(t)A_i(t) + \alpha_i S(t) \hat{Y}_i(t)] =  C_1 + \zeta^2 C_2  + \sum^n_{i=1} \hat{Q}_i(t)A_i(t) + \alpha_i S(t) \hat{Y}_i(t) \nonumber \\
& + \sum^n_{i=1} \zeta \alpha_i  S(t) \hat{Z}_i(t)-\sum^n_{i=1} \overline{S}_i(t)[\zeta \hat{Z}_i(t) + \hat{Q}_i(t) + \hat{Y}_i(t)] + \sum^n_{i=1} {D}^{'}_i(t) [V w_i - \zeta \hat{Z}_i(t) - \hat{Q}_i(t)]  \nonumber \\ &+ \sum^n_{i=1} [\hat{D}_i(t) - {D}^{'}_i(t)] [V w_i - \zeta \hat{Z}_i(t) - \hat{Q}_i(t)] \label{eq:lyapunov_eq1} \\
& {\leq}  C_1 + \zeta^2 C_2 + \sum^n_{i=1} [\zeta \alpha_i  S(t) \hat{Z}_i(t) + \hat{Q}_i(t)A_i(t) + \alpha_i S(t) \hat{Y}_i(t)] - \sum^n_{i=1} {S}^{*}_i(t)[\zeta \hat{Z}_i(t) + \hat{Q}_i(t) + \hat{Y}_i(t)] \nonumber \\
& + \sum^n_{i=1} {D}^{*}_i(t) [V w_i - \zeta \hat{Z}_i(t) - \hat{Q}_i(t)] + \sum^n_{i=1} [\hat{D}_i(t) - {D}^{'}_i(t)] [V w_i - \zeta \hat{Z}_i(t) - \hat{Q}_i(t)] \nonumber \\
&=  C_1 + \zeta^2 C_2 + \sum^n_{i=1} [\hat{D}_i(t) - {D}^{'}_i(t)] [V w_i - \zeta \hat{Z}_i(t) - \hat{Q}_i(t)] + \sum^n_{i=1} \zeta \hat{Z}_i(t)[\alpha_i  S(t) - {S}^{*}_i(t)- {D}^{*}_i(t)]  \nonumber \\
&+ \sum^n_{i=1} \hat{Q}_i(t)[A_i(t)- {S}^{*}_i(t) - {D}^{*}_i(t)] + \sum^n_{i=1} \hat{Y}_i(t)[\alpha_i S(t)- {S}^{*}_i(t)] + V \sum_{i=1}^{n} w_i {D}^{*}_i(t) \label{eq:lyapunov_eq2}
\end{align}
\hrulefill
\end{figure*}

We have established that for all $t \in [0, t_2)$, $\overline{Y}_i(t) \leq n(\beta + S^{max})  + S^{max}$. By repeatedly applying the above arguments to the interval $[t_2, \infty)$ and so on, we can show that the above-bound holds for all time slots $t \geq 1$. 
\end{IEEEproof}

 From Lemma~\ref{lem:y_queue_bound} and Proposition~\ref{prop:serv_const}, we can see that policy $\boldsymbol{\overline{\pi}}$ satisfies Constraint~\eqref{eq:service_constraint}. Further, Lemma~\ref{lem:data_queue_bound} presents an explicit upper bound that, in turn, establishes the stability of the data queues. Thus, policy $\boldsymbol{\overline{\pi}}$ meets the requirements to be a feasible solution of \eqref{eq:p1}.
 \hfill \IEEEQED

\subsection{Proof of Proposition~\ref{prop:performance} \label{sec:performance_proof}}

Let $\{\hat{Q}_i(t), i \in \mathcal{N} \}^{\infty}_{t=1}$, $\{\hat{Y}_i(t), i \in \mathcal{N} \}^{\infty}_{t=1}$, and $\{\hat{Z}_i(t), i \in \mathcal{N} \}^{\infty}_{t=1}$ denote the length of data, virtual and persistent queues under policy $\boldsymbol{\hat{\pi}}$. Next, we present a few  lemmas that help us prove Proposition~\ref{prop:performance}.

\begin{lem} \label{lem:qbounds_under_hat}
For all $ i \in \mathcal{N}, t \geq 1$, we have
\begin{enumerate}[(a)]
    \item $\hat{Q}_i(t)\leq Vw_i + A_i^{max}$
    \item $\hat{Z}_i(t) \leq Vw_i/\zeta + \zeta \alpha_i S^{max}$
    \item $\zeta \hat{Z}_i(t)+ \hat{Q}_i(t)\leq Vw_i+\zeta^2 \alpha_i S^{max}+A_i^{max}$
    \item $\hat{Y}_i(t)\leq  n(V+ \zeta^2 S^{max} + \max_{1 \leq i \leq n} A_i^{max}) + (n+1)S^{max}$
\end{enumerate}
\end{lem}
\begin{IEEEproof}
Similar to proof of Lemmas~\ref{lem:data_queue_bound}, \ref{lem:z_queue_bound}, \ref{lem:dataz_queue_bound} and \ref{lem:y_queue_bound}.
\end{IEEEproof}

\begin{lem} \label{lem:diff_bound}
For all $i \in \mathcal{N}$, 
$\frac{1}{V} \limsup_{T \to \infty} \sum^{T}_{t=1}(\hat{D}_i(t) -{D}^{'}_i(t))[Vw_i -\zeta \hat{Z}_i(t)-\hat{Q}_i(t)] \leq O \left( \frac{1}{V} \right) $, where $D^{'}_i(t) = \arg \min \limits_{x \in \{0,1,\ldots,D^{max}_i\}} (Vw_i - \zeta \hat{Z}_i(t) - \hat{Q}_i(t)) \cdot x$.
\end{lem}
\begin{IEEEproof}
When $Vw_i \geq \zeta \hat{Z}_i(t)+\hat{Q}_i(t)$, drop decisions \eqref{eq:p2_dropdecision} and \eqref{eq:p2_dropdecision_2} give us $\hat{D}_i(t) = {D}^{'}_i(t) = 0$. On the other hand, when  $Vw_i < \zeta \hat{Z}_i(t)+\hat{Q}_i(t)$, we have
\begin{align*}
 &(\hat{D}_i(t)-{D}^{'}_i(t))[Vw_i -\zeta \hat{Z}_i(t)-\hat{Q}_i(t)] \\&= ({D}^{'}_i(t)-\hat{D}_i(t))[ \zeta \hat{Z}_i(t)+\hat{Q}_i(t)-Vw_i] \\
&\overset{a}{\leq} D^{max}_i (\zeta^2 \alpha_i S^{max} + A^{max}_i) 
\end{align*}
where (a) follows due to ${D}^{'}_i(t) \in \{0,1,2,\ldots,D^{max}_i \}$ and Lemma~\ref{lem:qbounds_under_hat}. Consequently, we have
\begin{align*}
&\frac{1}{V} \limsup_{T \to \infty} \sum^{T}_{t=1}(\hat{D}_i(t) -{D}^{'}_i(t))[Vw_i -\zeta \hat{Z}_i(t)-\hat{Q}_i(t)] \\ 
& \leq {D^{max}_i (\zeta^2 \alpha_i S^{max} + A^{max}_i)}/{V} = O \left( {1}/{V} \right)
\end{align*}
\end{IEEEproof}

Let $\{{Q}^*_i(t), i \in \mathcal{N} \}^{\infty}_{t=1}$, $\{{Y}^{*}_i(t), i \in \mathcal{N} \}^{\infty}_{t=1}$, and $\{{Z}^{*}_i(t), i \in \mathcal{N} \}^{\infty}_{t=1}$ denote the length of the queues under optimal policy $\boldsymbol{\pi^{*}} =\{(S_i^*(t),D_i^*(t)) , i \in \mathcal{N} \}^{\infty}_{t=1}$.

\begin{lem} \label{lem:order_bound_1}
For all $i \in \mathcal{N}$, we have
$\frac{1}{V} \limsup_{T \to \infty} \frac{1}{T} \sum^T_{t=1} \hat{Q}_i(t)[A_i(t)-S^*_i(t)-D^*_i(t)] \leq O\left({1}/{V}\right)$.
\end{lem}
\begin{IEEEproof}
From \eqref{eq:dq}, we know that
$$A_i(t)-S^*_i(t)-D^*_i(t)  \leq {Q}^*_i(t+1)-{Q}^*_i(t)$$

\noindent
Consequently, we have
{
\begin{align*}
&\dfrac{1}{V} \limsup_{T \to \infty} \frac{1}{T} \sum^T_{t=1} \hat{Q}_i(t)[A_i(t)-S^*_i(t)-D^*_i(t)]\\
&{\leq} \dfrac{1}{V} \limsup_{T \to \infty} \frac{1}{T} \sum^T_{t=1}  Q^*_i(t)[\hat{Q}_i(t-1)-\hat{Q}_i(t)]\\
&+ \dfrac{1}{V} \limsup_{T \to \infty} \frac{ \hat{Q}_i(T)Q^*_i(T+1)}{T} \\
&\overset{a}{\leq} \dfrac{(S^{max}+D_i^{max})}{V} \limsup_{T \to \infty} \frac{1}{T} \sum^T_{t=1} Q^*_i(t) \\ 
&+ \dfrac{(Vw_i + A^{max}_i)}{V} \limsup_{T \to \infty} \frac{Q^*_i(T+1)}{T} \\
&\overset{b}{\leq} \dfrac{(S^{max}+D_i^{max})C_3}{V} = O\left(\dfrac{1}{V}\right)
\end{align*}}
where $C_3$ is a non-negative constant; (a) follows because $\hat{Q}_i(t-1)-\hat{Q}_i(t) \leq \overline{S}_i(t)+\hat{D}_i(t) \leq S^{max}+D_i^{max}$ and $\hat{Q}_i(T)\leq V w_i + A^{max}_i$; and (b) follows from rate stability of data queues under optimal policy $\boldsymbol{\pi^{*}}$.
    \end{IEEEproof}

\begin{lem} \label{lem:order_bound_2}
For all $i\in \mathcal{N}$, we have 
$\frac{1}{V} \limsup_{T \to \infty} \frac{1}{T} \sum^T_{t=1} \sum^N_{i=1}\hat{Z}_i(t)[\zeta \alpha_i S(t)-\zeta S^*_i(t)-\zeta D^*_i(t)] \leq O\left(\frac{1}{V}\right)+O\left(\frac{\epsilon}{V}\right)$.
\end{lem}
\begin{IEEEproof}
$\boldsymbol{\pi^{*}}$ is an optimal solution of \eqref{eq:p1}. Therefore, we have
$$ 0   \geq \limsup_{T \to \infty} \frac{1}{T} \sum^T_{t=1} \zeta ( \alpha_i S(t) - S^{*}_i(t)) 
\geq \limsup_{T \to \infty} \frac{1}{T} \sum^T_{t=1} \zeta ( \alpha_i S(t)  - S^*_i(t) - D^*_i(t))$$ 
Now, consider an auxiliary queue with the following evolution:
$Z_i^*(t+1)=[Z_i^*(t) - \zeta(S^*_i(t) + D^*_i(t))- \epsilon]^+ + \zeta \alpha_i S(t)$ where $\epsilon>0$.
Now, an application of  \cite[Lemma~2]{Shroff2012} with $\mu(t) = \zeta (S^*_i(t) + D^*_i(t)) + \epsilon  $ and $\lambda(t) = \zeta   \alpha_i S(t)$ tells us that the queue $Z^*_i(t)$ is strongly stable. 
\begin{align*}
   & \frac{1}{V} \limsup_{T \to \infty} \frac{1}{T} \sum^T_{t=1} \hat{Z}_i(t)[\zeta \alpha_i S(t) - \zeta S^*_i(t) - \zeta D^*_i(t)] \\
   & \leq  \frac{1}{V} \limsup_{T \to \infty} \frac{1}{T} \sum^T_{t=1}  \hat{Z}_i(t)[Z^*_i(t+1) -Z^*_i(t) + \epsilon]\\
   & \leq \frac{1}{V} \limsup_{T \to \infty} \frac{1}{T} \sum^T_{t=1}  {Z}^*_i(t)[\hat{Z}_i(t-1)-\hat{Z}_i(t)] \\
   &+ \dfrac{1}{V} \limsup_{T \to \infty} \frac{ \hat{Z}_i(T)Z^*_i(T+1)}{T}  +\dfrac{1}{V}  \limsup_{T \to \infty} \frac{1}{T} \sum^T_{t=1} \epsilon \hat{Z}_i(t) \\
   &\overset{a}{\leq} \dfrac{\zeta (S^{max}+D_i^{max})}{V} \limsup_{T \to \infty} \frac{1}{T} \sum^T_{t=1}  {Z}^*_i(t)\\
   & + \dfrac{(Vw_i/\zeta + \zeta \alpha_i S^{max})}{V} \limsup_{T \to \infty} \frac{Z^*_i(T+1)}{T} \\
   & + \dfrac{\epsilon}{V} \left( \dfrac{Vw_i}{\zeta} + \zeta \alpha_i S^{max} \right) \overset{b}{\leq} \dfrac{\zeta(S^{max}+D_i^{max})C_{\epsilon}}{V} \\
   & + \dfrac{\epsilon}{V} \left( \dfrac{Vw_i}{\zeta} + \dfrac{\zeta \alpha_i S^{max}}{V} \right) = O\left(\dfrac{1}{V}\right)+O\left(\epsilon\right)
\end{align*}
where $C_{\epsilon}$ is a non-negative constant that depends on $\epsilon$; (a) follows because $\hat{Z}_i(t-1)-\hat{Z}_i(t) \leq \zeta(\overline{S}_i(t)+\hat{D}_i(t)) \leq \zeta(S^{max}+D_i^{max})$ and $\hat{Z}_i(T)\leq Vw_i/\zeta + \zeta \alpha_i S^{max}$; and (b) follows from strong stability of $Z^*_i(t)$.
\end{IEEEproof}

\begin{lem} \label{lem:order_bound_3}
For all $i\in \mathcal{N}$, we have $\frac{1}{V} \limsup_{T \to \infty} \frac{1}{T} \sum^T_{t=1}Y_i(t)[\alpha_i S(t)-S^*_i(t)] \leq O\left(\frac{1}{V}\right)+O\left(\frac{\epsilon}{V}\right)$.
\end{lem}
\begin{IEEEproof}
Similar to the proof of Lemma~\ref{lem:order_bound_2}.
\end{IEEEproof}

Due to \eqref{eq:drift_p}, the drift-plus-penalty $\Delta \hat{L}(t) + V \hat{D}(t)$ under policy $\boldsymbol{\hat{\pi}}$ can be bounded as shown in \eqref{eq:lyapunov_eq1}, where ${D}^{'}_i(t) \in \{1,2,\ldots,D^{max}_i\}$ is the drop decision given by rule \eqref{eq:p2_dropdecision}, i.e., the one that minimizes the term ${D}_i(t)[Vw_i -\zeta \hat{Z}_i(t)-\hat{Q}_i(t)]$. Let $\boldsymbol{\pi^{*}} =\{(S_i^*(t),D_i^*(t)) , i \in \mathcal{N} \}^{\infty}_{t=1}$ be an optimal solution of \eqref{eq:p1}. We note that policy $\boldsymbol{\pi^{*}}$ may not be one-step optimal like policy $\boldsymbol{\overline{\pi}}$. Consequently, we can bound the R.H.S of \eqref{eq:lyapunov_eq1} as in \eqref{eq:lyapunov_eq2}. Summing \eqref{eq:lyapunov_eq2} over $t\in \{1 \dots T\}$, dividing throughout by $VT$, using $\hat{L}(T)/T \geq 0$ and taking limsup yields

\begin{align*}
&D^* \leq \hat{D} \leq D^* +\dfrac{C_1}{V}+ \dfrac{\zeta^2 C_2}{V}\\
&+\dfrac{1}{V} \limsup_{T \to \infty} \frac{1}{T} \sum^T_{t=1} \sum^n_{i=1} \zeta\hat{Z}_i(t)[\alpha_i S(t)-  S^*_i(t) - D^*_i(t)] \\
&+\dfrac{1}{V} \limsup_{T \to \infty} \frac{1}{T} \sum^T_{t=1} \sum^n_{i=1}\hat{Q}_i(t)[A_i(t)-S^*_i(t)-D^*_i(t)]\\
&+ \dfrac{1}{V} \limsup_{T \to \infty} \frac{1}{T} \sum^T_{t=1} \sum^n_{i=1}\hat{Y}_i(t)[\alpha_i S(t)-S^*_i(t)] \\
&+  \limsup_{T \to \infty} \frac{1}{T} \sum^T_{t=1} \sum^n_{i=1} \frac{[\hat{D}_i(t) - {D}^{'}_i(t)] [V w_i - \zeta \hat{Z}_i(t) - \hat{Q}_i(t)]}{V} \\
& \overset{(a)}{\leq} D^* + O\left(\frac{1}{V}\right) +O\left({\epsilon}\right) \overset{(b)}{=} D^* + O(\epsilon)
\end{align*}
\noindent
Here $\hat{D}=\limsup_{T \to \infty} \frac{1}{T} \sum^T_{t=1} \sum^n_{i=1} w_i \hat{D}_i(t)$, ${D}^*=\limsup_{T \to \infty} \frac{1}{T} \sum^T_{t=1} \sum^N_{i=1} w_i D_i^*(t)$ are the long-term weighted packet drops under polices $\boldsymbol{\hat{\pi}}$ and $\boldsymbol{{\pi}^{*}}$, respectively; (a) follows from Lemmas~\ref{lem:diff_bound}, \ref{lem:order_bound_1}, \ref{lem:order_bound_2} and \ref{lem:order_bound_3}; and
(b) follows by choosing a large enough value of $V$.

\subsection{Proof of Proposition~\ref{prop:delay_dynamic} \label{sec:delay_dynamic_proof}}
Let $M_i \geq 1$ be the worst-case delay (no. of slots) of flow $i$'s packets at gNodeB under policy $\boldsymbol{\hat{\pi}}$. Let us consider a packet that experiences a delay of $M_i$. Suppose this packet arrives in slot $t$, then it would remain in its data queue till the start of slot $t + M_i$. Then, for all time slots $\tau \in \{t+1,\dots ,t+M_i\}$, flow $i$'s data queue would have at-least one packet. Consequently, we have
$$\hat{Z}_i(\tau+1) \geq  \hat{Z}_i(\tau)+\zeta(\alpha_i S(\tau) - \overline{S}_i(\tau)- \hat{D}_i(\tau))$$

Summing the above inequality over $\tau \in \{t+1,\dots ,t+M_i\}$, rearranging the term and using the fact that $\hat{Z}_i(t)\geq 0$, we have
$$\sum_{\tau=t+1}^{t+M_i} \hspace{-3mm} \alpha_i S(t) \leq   \frac{\hat{Z}_i(t+M_i+1)}{\zeta} +  \sum_{\tau=t+1}^{t+M_i}  [\overline{S}_i(\tau)+\hat{D}_i(\tau)]$$

Since a packet from slot $t$ remains in its data queue till the start of slot $t + M_i$,  the number of packets served or dropped in slots $\{t+1,\ldots,t+M_i\}$ should be less than queue length at the beginning of slot $t+1$, i.e., $\sum \nolimits_{\tau=t+1}^{t+M_i}  [\overline{S}_i(\tau)+\hat{D}_i(\tau)] \leq \hat{Q}_i(t+1)$. Therefore, we have
\begin{align*}
\sum_{\tau=t+1}^{t+M_i} \alpha_i S(t)  &\leq \hat{Q}_i(t+1)+{\hat{Z}_i(t+M_i)}/{\zeta}\\
&\overset{(a)}{\leq}   V w_i + A_i^{max} + {V w_i}/{\zeta^2} +  \alpha_i S^{max} 
\end{align*}
where (a) follows from Lemma~\ref{lem:qbounds_under_hat}. Now, if $S_i(t) \geq S^{min} >0$, then we have
$\alpha_i S^{min}  M_i \leq V w_i + A_i^{max} + {V w_i}/{\zeta^2} +  \alpha_i S^{max}$.  

\hfill \IEEEQED

\bibliographystyle{IEEEtran}
\footnotesize
\bibliography{ref.bib}
\end{document}